\documentclass{iopart}
\usepackage[utf8]{inputenc}

\usepackage{color}
\usepackage{graphicx}
\usepackage{float}
\usepackage{iopams}
\usepackage{subfig}
\usepackage{acronym}
\usepackage[margin=10pt,font=small,format=hang]{caption}
\usepackage{multirow}
\usepackage{import}
\usepackage{lineno}
\usepackage{hyperref}
\usepackage{url}
\usepackage{xspace}
\usepackage{xcolor}
\usepackage{tabularx}
\usepackage{bigstrut}
\hypersetup{
    colorlinks=true,
    linkcolor=blue,
    filecolor=magenta,      
    urlcolor=cyan,
}

\DeclareGraphicsExtensions{%
    .pdf,.PDF,%
    .png,.PNG,%
    .jpg,.mps,.jpeg,.jbig2,.jb2,.JPG,.JPEG,.JBIG2,.JB2}

\def\mystrut{\vrule height 9.3pt depth 3.1pt width 0pt}
\definecolor{lightgray}{gray}{0.9} 

\begin{document}

\newcommand{\REF}[1]{\textcolor{red}{REFERENCE #1}}

\newcommand{\dmtdqvector}{{\sc DMT-DQ vector}\xspace}
\newcommand{\calibstate}{{\sc GDS-CALIB\_STATE vector}\xspace}
\newcommand{\dqr}{{\sc Data Quality Report}\xspace}
\newcommand{\idq}{i{\sc DQ}\xspace}
\newcommand{\ovl}{{\sc OVL}\xspace}
\newcommand{\qscan}{{\sc Q-scan}\xspace}
\newcommand{\hveto}{{\sc HVeto}\xspace}
\newcommand{\pointy}{{\sc pointy}\xspace}
\newcommand{\ldvw}{{\sc LIGO DataViewer Web}\xspace}
\newcommand{\lvalert}{{\sc LIGO-Virgo Alert System}\xspace}
\newcommand{\gracedb}{{\sc Gravitational-Wave Candidate Event Database}\xspace}
\newcommand{\gwpy}{{\sc GWpy}\xspace}
\newcommand{\gwdetchar}{{\sc GW-DetChar}\xspace}
\newcommand{\gwsumm}{{\sc GWSumm}\xspace}

\newcommand{\dd}[1]{\textcolor{violet}{#1}}

\newcommand{\lho}{{LIGO Hanford}\xspace}
\newcommand{\llo}{{LIGO Livingston}\xspace}
\newcommand{\LHO}{{LIGO Hanford}\xspace}
\acrodef{LHO}{\LHO}
\newcommand{\LLO}{{LIGO Livingston}\xspace}
\acrodef{LLO}{\LLO}
\newcommand{\virgo}{{Virgo}\xspace}

\newcommand{\pycbc}{{\sc PyCBC}\xspace}
\newcommand{\gstlal}{{\sc GstLAL}\xspace}
\newcommand{\cwb}{{\sc cWB}\xspace}

\newcommand{\gwcelery}{{\sc GWCelery}\xspace}

\acrodef{DetChar}{detector characterization}
\acrodef{IFO}{interferometer}

\acrodef{GW}{gravitational-wave}
\acrodef{LIGO}{Laser Interferometer Gravitational-wave Observatory}
\acrodef{BBH}{binary black hole}
\acrodef{O1}{first observing run}
\acrodef{O2}{second observing run}
\acrodef{O3}{third observing run}
\acrodef{O3a}{first half of the third observing run}
\acrodef{O3b}{second half of the third observing run}
\acrodef{O4}{fourth observing run}
\acrodef{BH}{black hole}
\acrodef{BBH}{binary black hole}
\acrodef{IMBH}{intermediate-mass black hole}
\acrodef{SNR}{signal-to-noise ratio}
\acrodef{BNS}{binary neutron star}
\acrodef{PSD}{power spectral density}
\acrodef{GR}{general relativity}
\acrodef{FAR}{false-alarm rate}
\acrodef{GCN}{the Gamma-ray Coordinates Network}
\acrodef{CBC}{compact binary coalescence}
\acrodef{VT}{volume-time}
\acrodef{ASD}{amplitude spectral density}
\acrodefplural{ASD}{amplitude spectral densities}
\acrodef{DAC}{digital-to-analog}
\acrodef{GWB}{gravitational-wave background}
\acrodef{DQ}{data quality}
\acrodef{RRT}{rapid response team}
\acrodef{GRB}{gamma-ray burst}

\acrodef{DARM}{differential arm readout measurement}
\acrodef{OSEM}{optical shadow sensors and magnetic actuator}
\acrodef{oplev}{optical lever}
\acrodef{ETM}{end test mass}
\acrodef{ETMY}{end test mass at the Y-end}
\acrodef{ETMX}{end test mass at the X-end}
\acrodef{AERM}{annular end reaction mass}
\acrodef{LSC}{length sensing and control}
\acrodef{PSL}{pre-stabilized laser}
\acrodef{FSS}{frequency stabilization system}
\acrodef{HAM3}{third horizontal access module}
\acrodef{L2}{penultimate}
\acrodef{L3}{third}
\acrodef{ESD}{electrostatic drive}
\acrodef{RC}{reaction chain}
\acrodef{ETG}{event trigger generator}
\acrodef{SQZ}{squeezer}
\acrodef{RMS}{root-mean-square}
\acrodef{RF}{radio frequency}
\acrodef{LED}{light emitting diode}
\acrodef{ASC}{alignment sensing and control}
\acrodef{TMS}{transmission motor stage}
\acrodef{OMC}{output mode cleaner}
\acrodef{DAQ}{data acquisition}
\acrodef{FFT}{fast Fourier transform}
\acrodef{ETG}{event trigger generator}
\acrodef{PRC}{Power Recycling Cavity}
\acrodef{SRC}{Signal Recycling Cavity}
\acrodef{AERM}{annular end reaction mass}
\acrodef{QPD}{quadrant photo-diode}

\acrodef{IGWN}{International Gravitational-wave Network}
\acrodef{X}{x}
\acrodef{$h(t)$}{gravitational-wave strain timeseries}
\acrodef{GWOSC}{Gravitational-wave Open Science Center}


\newcommand{\macro}[1]{\textcolor{black}{#1}}
\newcommand{\fixme}[1]{\textcolor{black}{#1}}

\newcommand{\Msun}{\ensuremath{\mathrm{M}_\odot}}

\newcommand{\EventJanuary}{\macro{GW170104}}
\newcommand{\EventJune}{\macro{GW170608}}
\newcommand{\LVTJuly}{\macro{LVT170729}}
\newcommand{\EventAugust}{\macro{GW170809}}
\newcommand{\EventTriple}{\macro{GW170814}}
\newcommand{\EventBNS}{\macro{GW170817}}
\newcommand{\LVTAugust}{\macro{LVT170823}}

\newcommand{\OTwoStart}{\macro{November 30, 2016}\xspace}
\newcommand{\OTwoEnd}{\macro{August 25, 2017}\xspace}
\newcommand{\OTwoObsTime}{\macro{August 25, 2017}\xspace}
\newcommand{\OTwoDuration}{\macro{268}\xspace}
\newcommand{\OTwoVirgoDuration}{\macro{25}\xspace}
\newcommand{\LHOOTwoRangeStart}{\macro{75~Mpc}\xspace}
\newcommand{\LHOOTwoRangeEnd}{\macro{65~Mpc}\xspace}
\newcommand{\LLOOTwoRangeStart}{\macro{80~Mpc}\xspace}
\newcommand{\LLOOTwoRangeEnd}{\macro{100~Mpc}\xspace}
\newcommand{\VirgoOTwoRange}{\macro{25~Mpc}\xspace}
\newcommand{\LHOOTwoRangeCleaningIncrease}{\macro{20\%}\xspace}

\newcommand{\OThreeStart}{\macro{April 1, 2019}\xspace}
\newcommand{\OThreeEnd}{\macro{March 27, 2020}\xspace}
\newcommand{\OThreeADuration}{\macro{183}\xspace}
\newcommand{\LHOOThreeARange}{\macro{108~Mpc}\xspace}
\newcommand{\LLOOThreeARange}{\macro{135~Mpc}\xspace}
\newcommand{\VirgoOThreeARange}{\macro{45~Mpc}\xspace}
\newcommand{\OThreeBDuration}{\macro{147}\xspace}
\newcommand{\OThreeBDurationLoss}{\macro{34}\xspace}
\newcommand{\LHOOThreeBRange}{\macro{113~Mpc}\xspace}
\newcommand{\LLOOThreeBRange}{\macro{131~Mpc}\xspace}
\newcommand{\VirgoOThreeBRange}{\macro{50~Mpc}\xspace}
\newcommand{\OTwoOThreeIncrease}{\macro{1.53-1.73}\xspace}
\newcommand{\LHOOTwoOThreeIncrease}{\macro{1.64}\xspace}
\newcommand{\LLOOTwoOThreeIncrease}{\macro{1.53}\xspace}
\newcommand{\VirgoOTwoOThreeIncrease}{\macro{1.73}\xspace}
\newcommand{\LLOOThreeBBHRange}{\macro{1425~Mpc}\xspace}
\newcommand{\LHOOThreeBBHRange}{\macro{1150~Mpc}\xspace}
\newcommand{\VirgoOThreeBBHRange}{\macro{525~Mpc}\xspace}
\newcommand{\OTwoOThreeIncreaseDutyCycle}{\macro{16\%}\xspace}
\newcommand{\OTwoOThreeIncreaseHLVDuration}{\macro{13}\xspace}

\newcommand{\LHODutyCycleOTwo}{\macro{65}}
\newcommand{\LHODutyCycleOThreeA}{\macro{71}}
\newcommand{\LHODutyCycleOThreeB}{\macro{79}}
\newcommand{\LHODutyCycleOThree}{\macro{75}}

\newcommand{\LLODutyCycleOTwo}{\macro{62}}
\newcommand{\LLODutyCycleOThreeA}{\macro{76}}
\newcommand{\LLODutyCycleOThreeB}{\macro{79}}
\newcommand{\LLODutyCycleOThree}{\macro{77}}

\newcommand{\VirgoDutyCycleOTwo}{\macro{85}}
\newcommand{\VirgoDutyCycleOThreeA}{\macro{76}}
\newcommand{\VirgoDutyCycleOThreeB}{\macro{76}}
\newcommand{\VirgoDutyCycleOThree}{\macro{76}}

\newcommand{\LHOpLLODutyCycleOTwo}{\macro{46}}
\newcommand{\LHOpLLODutyCycleOThreeA}{\macro{59}}
\newcommand{\LHOpLLODutyCycleOThreeB}{\macro{67}}
\newcommand{\LHOpLLODutyCycleOThree}{\macro{62}}

\newcommand{\LHOpLLOpVirgoDutyCycleOTwo}{\macro{63}}
\newcommand{\LHOpLLOpVirgoDutyCycleOThreeA}{\macro{44}}
\newcommand{\LHOpLLOpVirgoDutyCycleOThreeB}{\macro{51}}
\newcommand{\LHOpLLOpVirgoDutyCycleOThree}{\macro{47}}

\newcommand{\BLIPRATE}{\macro{2 per hour}\xspace}
\newcommand{\BLIPRATELLO}{\macro{4 per hour}\xspace}
\newcommand{\LHOLOUDGLITCHRATE}{\macro{3.5 per hour}\xspace}
\newcommand{\LLOLOUDGLITCHRATE}{\macro{3.3 per hour}\xspace}

\newcommand{\LHOCATONEOTWO}{\macro{1.93}}
\newcommand{\LLOCATONEOTWO}{\macro{0.49}}
\newcommand{\LHOCATTWOCBCOTWO}{\macro{0.26}}
\newcommand{\LLOCATTWOCBCOTWO}{\macro{0.16}}
\newcommand{\LHOCATTWOBURSTOTWO}{\macro{0.35}}
\newcommand{\LLOCATTWOBURSTOTWO}{\macro{0.26}}
\newcommand{\LHOCATTHREEBURSTOTWO}{\macro{0.61}}
\newcommand{\LLOCATTHREEBURSTOTWO}{\macro{0.33}}

\newcommand{\LHOCATONEOTHREEA}{\macro{0.27}}
\newcommand{\LLOCATONEOTHREEA}{\macro{0.08}}
\newcommand{\LHOCATTWOCBCOTHREEA}{\macro{0.37}}
\newcommand{\LLOCATTWOCBCOTHREEA}{\macro{0.10}}
\newcommand{\LHOCATTWOBURSTOTHREEA}{\macro{0.83}}
\newcommand{\LLOCATTWOBURSTOTHREEA}{\macro{0.64}}
\newcommand{\LHOCATTHREEBURSTOTHREEA}{\macro{0.19}}
\newcommand{\LLOCATTHREEBURSTOTHREEA}{\macro{0.15}}

\newcommand{\LHOCATONEOTHREEB}{\macro{0.30}}
\newcommand{\LLOCATONEOTHREEB}{\macro{1.68}}
\newcommand{\LHOCATTWOCBCOTHREEB}{\macro{0.02}}
\newcommand{\LLOCATTWOCBCOTHREEB}{\macro{0.28}}
\newcommand{\LHOCATTWOBURSTOTHREEB}{\macro{0.52}}
\newcommand{\LLOCATTWOBURSTOTHREEB}{\macro{0.50}}
\newcommand{\LHOCATTHREEBURSTOTHREEB}{\macro{0.41}}
\newcommand{\LLOCATTHREEBURSTOTHREEB}{\macro{0.17}}

\newcommand{\GWTCTWOMITIGATION}{\macro{10}\xspace}

\newcommand{\RESPONSETIME}{\macro{35 minutes}\xspace}

\newcommand{\DANGERPVALUE}{\macro{$2 \times 10^{-4}$}\xspace}
\newcommand{\UNSAFECHANNELSPERDETECTOR}{\macro{55}\xspace}
\newcommand{\TOTALCHANNELSANALYZED}{\macro{3000}\xspace}
\newcommand{\PERCENTOFCHANNELSUNSAFE}{\macro{2}\xspace}

\title[]{LIGO Detector Characterization in the Second and Third Observing Runs}
\author{%
D~Davis$^{1}$,  
J~S~Areeda$^{2}$,  
B~K~Berger$^{3}$,  
R~Bruntz$^{4}$,  
A~Effler$^{5}$,  
R~C~Essick$^{6}$,  
R~P~Fisher$^{4}$,  
P~Godwin$^{7}$,  
E~Goetz$^{8,9,10}$,  
A~F~Helmling-Cornell$^{11}$
% ...
}%
\medskip
\address {$^{1}$LIGO, California Institute of Technology, Pasadena, CA 91125, USA }
\address {$^{2}$California State University Fullerton, Fullerton, CA 92831, USA }
\address {$^{3}$Stanford University, Stanford, CA 94305, USA }
\address {$^{4}$Christopher Newport University, Newport News, VA 23606, USA }
\address {$^{5}$LIGO Livingston Observatory, Livingston, LA 70754, USA }
\address {$^{6}$University of Chicago, Chicago, IL 60637, USA }
\address {$^{7}$The Pennsylvania State University, University Park, PA 16802, USA }
\address {$^{8}$University of British Columbia, Vancouver, BC V6T 1Z4, Canada }
\address {$^{9}$Missouri University of Science and Technology, Rolla, MO 65409, USA }
\address {$^{10}$Louisiana State University, Baton Rouge, LA 70803, USA }
\address {$^{11}$University of Oregon, Eugene, OR 97403, USA }
% ...

\date{\today}

\begin{abstract}
The characterization of the Advanced LIGO detectors in the second
and third observing runs has increased
the sensitivity of the instruments, allowing for 
a higher number of detectable gravitational-wave signals,
and provided confirmation of all observed 
gravitational-wave events. 
In this work, we present the methods
used to characterize the LIGO detectors and curate 
the publicly available datasets, including the LIGO strain data
and data quality products. 
We describe the essential role of these datasets in LIGO-Virgo Collaboration 
analyses of gravitational-waves from 
both transient and persistent sources and include details on the 
provenance of these datasets in order to support analyses of 
LIGO data by the broader community. 
Finally, we explain anticipated changes in the role of detector characterization 
and current efforts to prepare for the high rate of gravitational-wave
alerts and events in future observing runs. 
\end{abstract}

\maketitle

\section{Introduction}\label{s:intro}

The Laser Interferometer Gravitational-wave 
Observatory (LIGO)~\cite{aLIGO} 
and Virgo~\cite{TheVirgo:2014hva} are the most sensitive facilities for 
the direct detection of gravitational waves. 
They have been observing the gravitational wave sky 
in their advanced configuration since 2015 and 
in a total three observing runs so far. 
Characterization of the LIGO detectors enabled 
and enhanced the discoveries reported by 
LIGO-Virgo in their \ac{O2} and \ac{O3}.
The two LIGO detectors participated in \ac{O2} 
from \OTwoStart to \OTwoEnd, and the Virgo detector 
joined for the last 25 days of the run. 
All three LIGO and Virgo detectors took data during \ac{O3},
from \OThreeStart to \OThreeEnd.
The LIGO-Virgo Collaboration has since 
reported the confident detection of gravitational wave signals from 
seven black hole mergers and one binary neutron star merger 
during O2 in GWTC-1~\cite{LIGOScientific:2018mvr} 
and 39 detections of black hole and neutron star mergers
during the first half of \ac{O3} in GWTC-2~\cite{GWTC-2}.

The two US-based LIGO detectors are dual-recycled 
Michelson interferometers with 4~km Fabry-Perot arm cavities. The
LIGO detectors are designed to sense extremely small fluctuations
in spacetime induced by passing gravitational waves~\cite{aLIGO}.
\ac{LHO} is located in Hanford, Washington, and \ac{LLO} 
is located in Livingston, Louisiana. 
During \ac{O2} and \ac{O3}, some differences in 
configuration between the \ac{LHO} and \ac{LLO} 
instruments resulted in differences in 
technical noise sources contributing to effective gravitational wave strain 
noise between the two detectors, 
as reported in ~\cite{Driggers:2018gii,Buikema:2020dlj}. 

LIGO detector data is a gravitational-wave strain time series 
that is rich with noise artifacts.
Often the noise is dominated by fundamentally limiting noise 
sources~\cite{aLIGO},  
causing it to appear Gaussian and stationary over limited 
time scales and frequency ranges.  
The sensitivity of the detectors as measured by the amplitude
of these noise sources, combined with the coincident uptime with multiple detectors observing, 
are key metrics of detector performance.
However, LIGO data also contains a high rate of transient noise
artifacts, or \textit{glitches}, that contribute to the noise background 
of searches for gravitational waves by mimicking the behavior of
true astrophysical signals. Glitches can also 
overlap with signals, as reported in~\cite{GW170817}, 
and confuse source property 
estimation of even confidently detected signals unless properly 
mitigated~\cite{Pankow:2018qpo, Powell:2018csz, Cornish:2020dwh}.
LIGO data also contains strong nearly sinusoidal features, 
or \textit{lines}, that inhibit searches for long duration sources of 
gravitational waves, as described in~\cite{Covas:2018}.
Additionally, LIGO detector data 
exhibits slow changes to the characteristics of the noise due to 
complex interactions between the 
detectors and their local environment. 

In order to address these features of the data that differ from the 
output of an idealized gravitational-wave interferometer, 
the LIGO detectors and data are closely monitored 
before and during observing runs
using a large number of additional data streams 
(referred to as \emph{auxiliary channels}), that 
include sensors of the environment surrounding the detectors
and measurements of the detector control systems.
These efforts to understand and mitigate these sources of noise, both in 
the instrument and the data are collectively referred to as 
``detector characterization''.  
Detector characterization is an essential component of 
improving the performance of the LIGO detectors
and the detection of gravitational wave 
events~\cite{Nuttall:2015dqa,GW150914_detchar}.

\begin{figure}[tb]
    \centering
    \includegraphics[width=\textwidth]{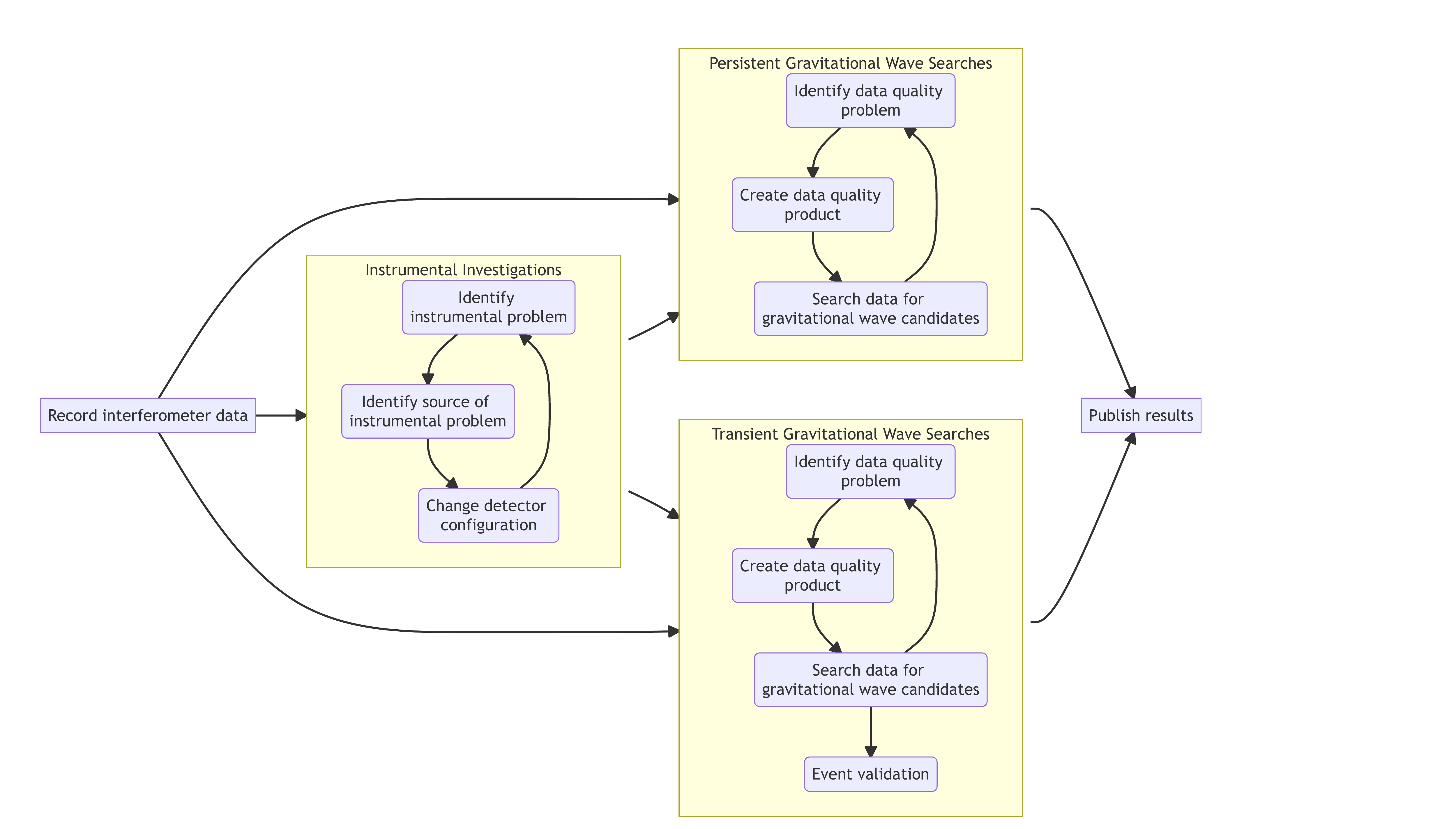}
    \caption{Flowchart outlining the the flow of information through 
    various detector characterization analyses.
    The main areas of focus for detector characterization are instrumental 
    investigations and data quality investigations.
    These data quality investigations are further broken down based on the 
    type of gravitational-wave analysis of interest. 
    Note that this flow chart only represents a high-level overview of the work
    both in this paper and in detector characterization research in general. 
    Additional connections between detector characterization 
    and the operation of the LIGO detectors and the analysis of LIGO data are
    also not shown in this flowchart. 
    The items in this flowchart are further explained 
    individually later in the text.}
    \label{figure:flow}
\end{figure}

Detector characterization research in LIGO focuses on 
both investigations of the instruments in order to improve
the future performance of the detectors and
investigations of data quality in order to improve the performance
of gravitational wave analyses. 
These data quality investigations focus on seperate types 
of data quality issues depedning on the gravitational-wave analysis
of interest. 
The majority of data quality issues that impact searches
for short duration, transient gravitational-wave signals
do not impact searches for long duration, persistent
gravitational-wave signals, and vice versa.  
An overview of the the connections between the multiple
aspects of dectector characterization is shown in \fref{figure:flow}. 

LIGO data 
is publicly distributed via the \ac{GWOSC}~\cite{Abbott:2019ebz}.
Currently available data includes
gravitational wave strain data during periods the individual detectors
were observing in the first two observing runs
and data quality information used
in LIGO analyses~\cite{gwosc_o1_strain,gwosc_o2_strain}.
LIGO data from the third observing run is planned to be released 
in six month periods, 18 months after the start of each observing
period~\cite{data_management}.
In addition to these bulk data releases,
data nearby all detected gravitational-wave events is released
via \ac{GWOSC} at the time of publication.
Data from a subset of auxiliary channels is currently available
for a three hour period around a single event~\cite{gwosc_aux_data}. 

In this paper, we report the results of detector characterization 
methods applied to LIGO detector data from \ac{O2} and \ac{O3} 
to improve the performance of the 
detectors and astrophysical analyses.
In \sref{s:O2dataset} we summarize the LIGO \ac{O2} and \ac{O3}  
data sets as reported in~\cite{LIGOScientific:2018mvr,GWTC-2}.
In \sref{s:computing} we describe major tools and infrastructure
employed for LIGO detector characterization during these observing runs.
In \sref{s:noiseinvestigations} we outline work that improved the performance
of the LIGO detectors by characterizing 
and mitigating sources of instrumental noise.
In \sref{s:transientDQ} we summarize the methodology of LIGO
data quality products employed by transient gravitational wave searches using
\ac{O2} and \ac{O3} data as well as methods and 
procedures applied to LIGO detector data to validate transient event candidates. 
In \sref{s:longdurDQ} we describe data quality investigations
and products used by searches for gravitational waves from 
persistent sources.
We conclude in \sref{s:O4expectations} with an overview of 
future work, including automation efforts designed to cope with the
significantly higher sensitivity and expected event rate during future observing runs.


\section{The O2 and O3 data sets}\label{s:O2dataset}
The \ac{O2} period spanned \OTwoDuration calendar days, with the LIGO 
detectors participating for the entire period. Virgo, however, joined for the 
last \OTwoVirgoDuration 
days. There were two scheduled breaks over the observing run; the 2016 
end-of-year holidays and a few weeks in May 2017, which was used to make 
improvements to each of the LIGO detectors.

\begin{figure}[ht]
\includegraphics[width=\textwidth]{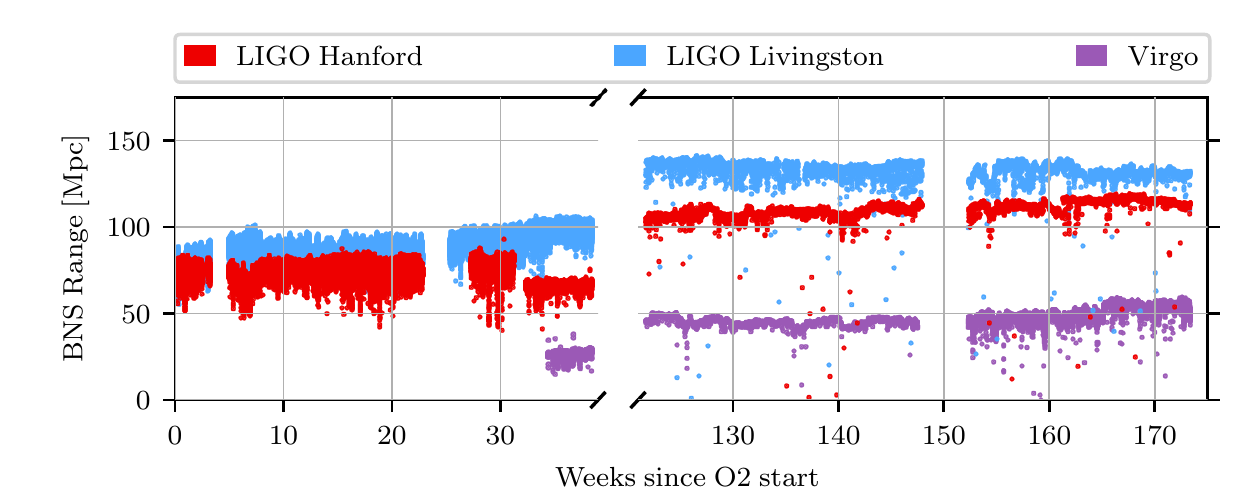}
\includegraphics[width=\textwidth]{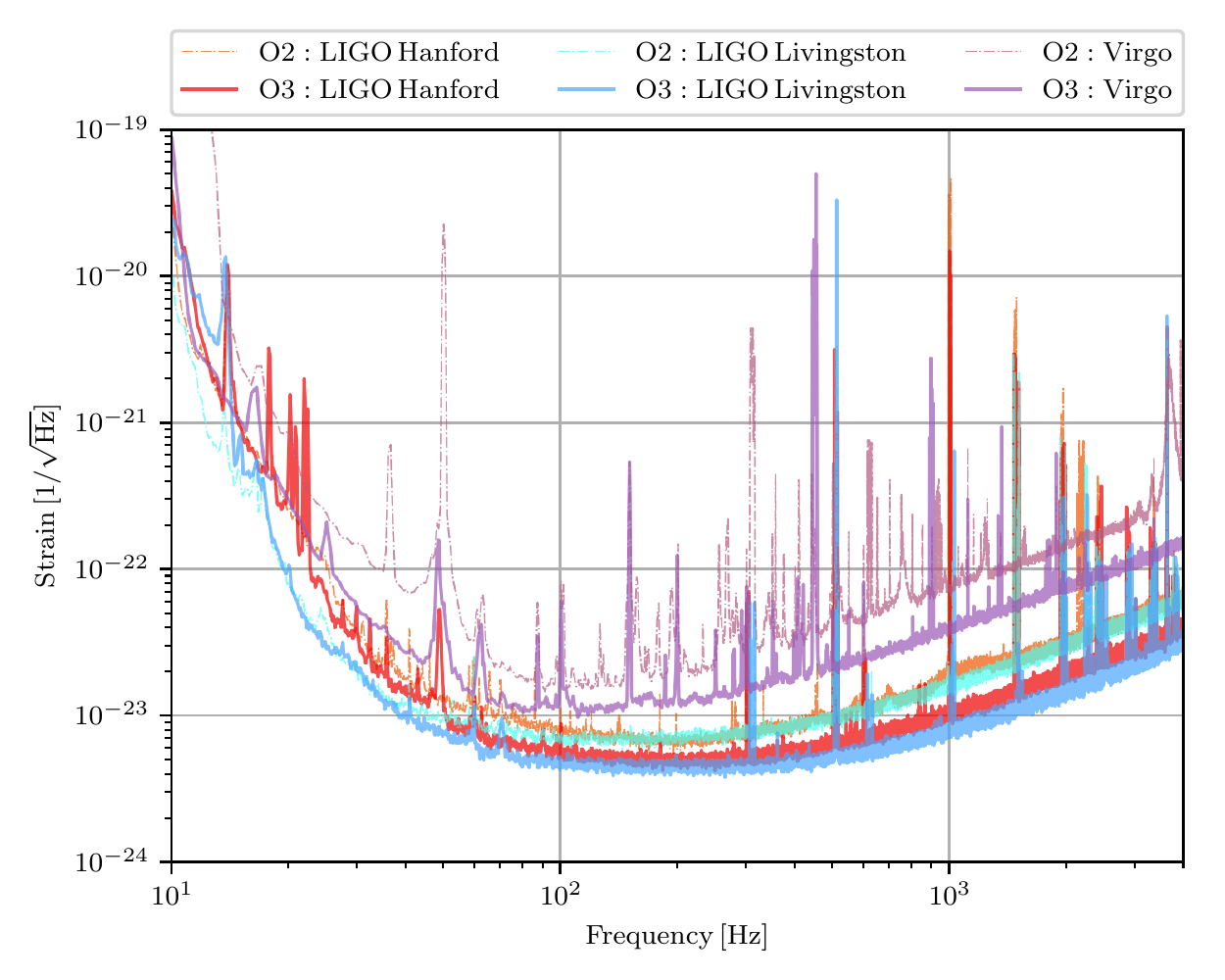}
\caption{Top: Binary Neutron Star (BNS) range evolution of the LIGO and Virgo 
detectors from the start of O2 in November 2017 to the end of O3 in March 2020. 
The broken axes remove the time between each observing run. Bottom: 
Representative amplitude spectral density of the three detectors' strain 
sensitivity in each observing run. The O3 spectra shown are taken from O3a.}
\label{figure:range_asd}
\end{figure}

One way in which we measure sensitivity is by the binary neutron star inspiral 
range; this range is the distance at which a gravitational-wave signal from a 
the merger of two $1.4 M_\odot$ neutron stars
would be detected above a \ac{SNR} of 8,
averaged over all possible sky locations and inclinations
without considering cosmological corrections. The \ac{LLO} 
detector started 
\ac{O2} observing around \LLOOTwoRangeStart, and became steadily more sensitive as 
\ac{O2} progressed, reaching
\LLOOTwoRangeEnd. The \ac{LHO} detector's sensitivity was around 
\LHOOTwoRangeStart at the start
of the observing run. It however, suffered a sudden drop in sensitivity on 6th 
July 2017 due to a 5.8 magnitude earthquake in Montana, finishing the run 
around \LHOOTwoRangeEnd. 
Virgo held a steady sensitivity around \VirgoOTwoRange for its 
\OTwoVirgoDuration-day observing period. 
This information is illustrated in \fref{figure:range_asd}. 

The \ac{O3} observing run was split into two periods, separated by the month of 
October 2019 to make stability 
improvements to all three detectors. 
The \ac{O3a} lasted for \OThreeADuration days with the \ac{LLO}, 
\ac{LHO} and Virgo detectors having a median range of 
\LLOOThreeARange, \LHOOThreeARange 
and \VirgoOThreeARange respectively. Due to the improvements made to the 
interferometers~\cite{GWTC-2} between \ac{O2} and \ac{O3}, 
the sensitivity of the detectors increased by a factor of 
\LLOOTwoOThreeIncrease for \ac{LLO}, \LHOOTwoOThreeIncrease for 
\ac{LHO} and \VirgoOTwoOThreeIncrease for Virgo. During \ac{O3b} the 
sensitivity of the detectors were similar to \ac{O3a}, with \ac{LLO}, 
\ac{LHO} and Virgo each having a median range of \LLOOThreeBRange, 
\LHOOThreeBRange and \VirgoOThreeBRange. \ac{O3b} lasted 
\OThreeBDuration days, some \OThreeBDurationLoss days less 
than was originally intended. If we instead consider the
sensitivity of the detectors to the inspial of two black holes each with a mass
of $30 M_\odot$, the ranges become approximately \LLOOThreeBBHRange,
\LHOOThreeBBHRange and \VirgoOThreeBBHRange for \ac{LLO}, \ac{LHO} and Virgo
respectively, throughout O3.

\Fref{figure:range_asd} shows the typical amplitude spectral 
density of the strain noise for each detector over \ac{O2} and \ac{O3}. 
The duty cycle of each detector defines the amount of science quality data 
taken over a period of time. There are a number of factors which affect the 
duty cycle, such as the environment (e.g., weather), detector hardware (e.g., 
malfunctioning instrument components) and periods of commissioning. 
\Tref{tab:duty} highlights the duty cycle of each of the detectors in 
\ac{O2} and \ac{O3}. 
We also give the coincident duty cycle of the LIGO detectors, 
as well as the triple coincident time. There is a marked improvement in 
the stability of the LIGO detectors between \ac{O2} and \ac{O3}, 
with coincident science 
quality time increasing by some \OTwoOThreeIncreaseDutyCycle. 
Although the Virgo duty cycle appears to 
decrease between observing runs, it should be highlighted that the \ac{O2} 
duty cycle  includes livetime that is about \OTwoOThreeIncreaseHLVDuration 
times less than in \ac{O3}. 
As well, the \OTwoVirgoDuration-day 
\ac{O2} time that Virgo was observing for included optimal environmental 
conditions. 
\\\\
\begin{table*}[t]
\begin{tabularx}{\textwidth}{l@{\extracolsep{\fill}}rrrr}
\multicolumn{5}{c}{\textbf{Duty Cycle}} \\
\hline
\textbf{Detector} & \textbf{O2} & \textbf{O3a} & \textbf{O3b} & \textbf{O3} \\
\hline

\makebox[0pt][l]{\fboxsep0pt\colorbox{lightgray}{\mystrut\hspace*{0.97\linewidth
}}}LHO & \LHODutyCycleOTwo \% & \LHODutyCycleOThreeA \% &
\LHODutyCycleOThreeB \% & \LHODutyCycleOThree \% \\
        LLO & \LLODutyCycleOTwo \% & \LLODutyCycleOThreeA \% &
\LLODutyCycleOThreeB \% & \LLODutyCycleOThree \% \\

\makebox[0pt][l]{\fboxsep0pt\colorbox{lightgray}{\mystrut\hspace*{0.97\linewidth
}}}Virgo & \VirgoDutyCycleOTwo \% & \VirgoDutyCycleOThreeA \% &
\VirgoDutyCycleOThreeB \% & \VirgoDutyCycleOThree \% \\
        LHO+LLO & \LHOpLLODutyCycleOTwo \% & \LHOpLLODutyCycleOThreeA \% &
\LHOpLLODutyCycleOThreeB \% & \LHOpLLODutyCycleOThree \% \\

\makebox[0pt][l]{\fboxsep0pt\colorbox{lightgray}{\mystrut\hspace*{0.97\linewidth
}}}LHO+LLO+Virgo & \LHOpLLOpVirgoDutyCycleOTwo \%  & \LHOpLLOpVirgoDutyCycleOThreeA \% &
\LHOpLLOpVirgoDutyCycleOThreeB \% & \LHOpLLOpVirgoDutyCycleOThree \% \\

\hline
\end{tabularx}
        \caption{The duty cycle (round to the nearest integer) of each of the
detectors, LIGO Hanford (LHO), LIGO Livingston (LLO) and Virgo, and
combinations, over the second (O2) and third observing run (O3). The O3 numbers 
are a combination of O3a and O3b. The Virgo and three detector duty cycles for 
O2 are normalized to the \OTwoVirgoDuration days that Virgo joined the 
observing run.}
\label{tab:duty}
\end{table*}

In both observing runs, we used auxiliary channels that recorded the source of
the instrumental noise (referred to as a ``witness'') to measure sources of 
noise that limit detector sensitivity. Using these measurements, we were able 
to linearly subtract this noise from the data.
During \ac{O2} a pipeline was developed to do this subtraction, which is easily
adaptable to target new sources of noise as they
arise~\cite{Driggers:2018gii, Davis:2019}. For both LIGO detectors this was
used to target narrow line features, such as the calibration lines
and 60~Hz and its harmonic frequencies. At
\ac{LHO}, however, there was an additional source of broadband noise known 
as
jitter noise. This form of noise was related to the jitter of the 
pre-stabilized laser beam in angle and size. This was only present at \ac{LHO} 
due to
different configurations between the two LIGO detectors. By subtracting this
form of noise below 1000~Hz, \ac{LHO} saw an average increase in
range, over \ac{O2}, of \LHOOTwoRangeCleaningIncrease~\cite{Davis:2019}. 
The \ac{LLO} detector saw no appreciable increase in its range. 

In \ac{O3} issues of jitter noise had been resolved, and so the same level of 
data
cleaning was not necessary. The removal of the calibration lines and noise from
their harmonics were subtracted from data as part of the calibration procedure.
For a subset of gravitational wave events detected during 
\ac{O3}, additional data 
cleaning was performed to remove noise contributions due to non-stationary 
couplings of the power mains~\cite{Vajenta:2020ml}.

As the interferometers are upgraded and improved, and more hardware goes into 
the interferometers, inevitably, there are more data quality issues that arise. 
For instance, \ac{O3} saw the installation of squeezed light 
sources~\cite{Barsotti_2018, Snabel:2010sq}. This is an additional system that 
could and did introduce additional noise into the \ac{O3} data. As the 
detectors become more stable, not only does their duty cycle improve, the 
increased observing time allows for more data quality issues to occur. These 
data quality issues are discussed in detail in the remainder of this paper. 

\section{Computing and Software}\label{s:computing}

Because ground-based gravitational-wave detectors are subject to a wide range
of environmental noise source and continual upgrades during an observing run,
and because new technologies periodically
emerge to improve sensitivity across the full frequency range, the detectors
themselves are continually evolving. This presents an endless challenge to
any effort to characterize noise in the detectors, as the source, shape, rate,
and intensity of various noise sources is constantly changing. In this section
we will outline the multiple computational solutions which have emerged to
help combat this problem, focusing on the types of analyses that each software
application is suited to. In doing so, we will build important context for
the methods and results presented later in the paper.
This section is not meant as an exhaustive list of all analysis tools that 
are used in detector characterization studies, 
but it serves to give a broad example and context for results 
discussed in this article.

\subsection{Signal processing tools}
\label{packages}

A number of open source computing projects are developed and maintained to
enable data analysis for LIGO detector characterization. These
tools are published through widely used version control platforms and
delivered to users through the \ac{IGWN} Conda Distribution~\cite{igwn_conda}.
Unless otherwise noted, they are written entirely in Python~\cite{python} and are
available under terms of the GNU General Public License, version 3.0.0.

These signal processing tools are designed to both process raw timeseries 
that are generated by the wide variety of data streams at each observatory, 
as well as other pre-processed data. 
One of the main types of pre-processed data types that these tools
are designed to ingest are ``triggers'' created by 
``event trigger generators.''
A wide variety of event trigger generators exist, but in general, are designed to 
find excess power in data streams. 
These excess power bursts are considered triggers. 
While some event trigger generators are designed to identify generic bursts 
of excess power using wavelets, other event trigger generators 
use waveform templates from general relativity 
to identify triggers that are consistent with a particular
gravitational-wave source. 

\subsubsection{GWpy}

The central signal processing and data visualization engine used to prepare
most figures in this paper is \gwpy~\cite{gwpy,gwpy_ascl}, a Python package
for studying data from gravitational-wave detectors. This package is designed
with an extensive set of features for manipulating data in both the time and
frequency domain, including:

\begin{enumerate}
\item Native memory-optimized Python classes for \texttt{TimeSeries} and
      \texttt{FrequencySeries} objects

\item Robust data input/output capabilities, including support for multiple
      file formats as well as time optimization through multithreading

\item Custom filtering applications, digital filter designers, and convolution
      algorithms tailored for \ac{IGWN} data

\item An implementation of both the fast Fourier transform and the multi-Q
      transform (see \sref{qscans}) for timeseries data

\item A \texttt{Table} class primarily designed for analyzing the output of
      various event trigger generators

\item Publication-quality visualization methods that are fine-tuned for every
      data product while remaining highly customizable
\end{enumerate}

While \gwpy is aimed at individual users and is relatively general-purpose, a
number of other packages with narrower scope are also derived from it, as
described below.

\subsubsection{GW-DetChar}

An extension of \gwpy with specific applications to \ac{IGWN} (especially LIGO)
detector characterization tasks is available in the \gwdetchar software package~\cite{gwdetchar}.
This codebase contains a number of user modules and scripts 
that are able to identify and analyze known
classes of glitches, such as optical scattering~\cite{Accadia:2010zzb},
as well as more general noise hunting algorithms such as Lasso regression
\cite{Walker:2018ylg}. For each tool in the package, the primary data product
is a single webpage with responsive design features~\cite{gwbootstrap} which
can be used to record and easily interpret results (see \sref{summary_pages}).

\subsubsection{Omega scans}
\label{qscans}

A particular data visualization submodule within \gwdetchar is the
\texttt{gwdetchar-omega} command-line tool, so named because it is a Python
implementation of a legacy unmodeled transient search pipeline called Omega
\cite{Chatterji:2004qg,Chatterji_thesis,Rollins10Thesis}. 
In a detector characterization context this tool is
used to identify and visualize the time-frequency morphology of various
sources of transient noise. Its primary data product is referred to as
an omega scan, which consists of a raw multi-Q transform.
The multi-Q transform consists of multiple spectrograms
that tile the time-frequency plane using tiles with constant ratio
of duration to bandwidth (known as the quality factor, or Q). 
This spectrogram tiling is referred to as the 
constant-Q transform~\cite{Chatterji:2004qg}.
The choice of Q that contains the maximum energy in a single
time-frequency tile is then chosen as the optimal value of Q. 
This optimized raw constant-Q transform is then interpolated, providing a qualitative
high-resolution image of signal energy as a function of time and frequency.

Through configuration files, users have the ability to analyze an arbitrary
number of data streams over the same time range, which makes the omega scan a
powerful tool in tracing the propagation of a glitch throughout interferometer
subsystems. To assist with this, \texttt{gwdetchar-omega} can optionally
cross-correlate every successive independent data stream with the 
signal, then display a tabulated ranking of the most highly correlated
channels. Alternatively, the omega scan can be used to process gravitational wave strain streams from an
arbitrary number of interferometers for a quick visual comparison of the signal
morphology in each stream over a fixed time interval.

\subsubsection{Omicron}
\label{omicron}

The primary event trigger generator for detector characterization studies is an unmodeled
transient detection pipeline called Omicron
\cite{Robinet:2020lbf}.
The Omicron pipeline broadly performs a multi-Q
transform given some data stream, then searches for significant clusters of
tiles in time-frequency space, optimizing over the quality factor. For each
LIGO detector, Omicron is run on the gravitational wave strain channel and
a collection of some 800-900 separate
channels representing interferometer subsystems, with triggers stored in a
central location from their on-site computing clusters. To ensure stability
of trigger production, the workflow is managed by a Python package called
\textsc{pyomicron}~\cite{pyomicron} and most channels have triggers available
with modest 1 hour latency.

\subsubsection{Hierarchical Veto}
\label{hveto}

While omega scans can be used to identify correlations between an arbitrary 
number of data streams, 
such analyses tend to be confined to a narrow window of time ($\sim 1$ sec)
to understand the origin of a specific transient glitch. 
On the other hand,
it is well worth understanding broader, longer-term correlations that may
exist over the course of hours or days and to assign a statistical signifcance
to the identified correlation. 
This is the purview of \hveto
\cite{Smith:2011an,hveto}, a companion to \gwdetchar that analyzes concurrent
patterns between clusters of event triggers above a fixed \ac{SNR} threshold
in multiple data streams.

\hveto correlation
searches are used to identify potentially statistically significant coincidences
between  Omicron~\cite{Robinet:2020lbf} triggers in the gravitational wave strain channel
and other auxiliary channels.  
The significance is calculated
as the probability of the number of observed coincidences divided by the number
expected (See~\cite{Smith:2011an}). 
\hveto is used multiple times per day to correlate glitches with auxiliary
channels that may interfere with the identification of gravitational waves.
When a significant association is found, \hveto analyzes the effect of removing
the time segments containing the associated glitches from the analysis
before proceeding to other data streams in a
hierarchical fashion. 
With each successive round of vetoes, a list of
statistically significant correlations emerges, 
ranked in descending order of significance.
To better understand the cause of the identified correlations, 
omega scans of a subset of the time periods removed in each round
are generated for visual inspection. 

\subsection{Web-based services}

By contrast with the software packages described in \sref{packages},
the following services are maintained as broad signal processing platforms
primarily accessible to the end user through the Internet, with application
programming interfaces (APIs) available on the command-line and through any
computing environment that supports Python. Like most LIGO and \ac{IGWN} 
web-based
services, they utilize Shibboleth Single Sign-on~\cite{shibb} for user
authentication to ensure the security of proprietary datasets.

\subsubsection{DQSEGDB} \label{ss:dqsegdb}
Many detector characterization tasks and pipelines designed to search for 
gravitational wave signals rely on data quality ``flags.'' 
Data quality flags store metadata for measured or derived states within each 
interferometer, its subsystems, and various components.
Data quality flags are used at all current gravitational wave observatories. 

These data quality flags are stored in the 
\ac{IGWN} Data Quality Segment Database (DQSEGDB)~\cite{Fisher:2020pnr}.  
For each flag, the database tracks spans
of time (called \emph{segments}) over which the flag’s on/off truth value is
known. 
The start and end time of each segment is defined using units of integer seconds.
In particular, each interferometer's ``observing mode'' flag indicates
segments over which that interferometer was both locked and taking
science-quality data that is flagged by interferometer operators as intended
for gravitational wave searches.  This flag is used by all
downstream analysis pipelines to distinguish spans of time that can and cannot
be analyzed.  Other flags can be used to reject (or \emph{veto}) otherwise
usable segments in which the data stream contains well-understood artifacts,
such as glitches with a known cause, or planned
injections of artificial test signals.

\subsubsection{Detector characterization summary pages}
\label{summary_pages}

For the convenience of LIGO commissioners, detector engineers, and data 
analysts,
an extensive suite of detector characterization summary pages is provided~\cite{ligo_summary}
which offer automated daily analyses of the primary gravitational wave strain data as well 
as
sundry interferometer subsystems. These pages are available to anyone with
federated credentials on the \texttt{LIGO.ORG} domain, and are batch-generated
on dedicated hardware with modest (0.5-1 hour) latency. 
The detector characterization summary pages are one of the main tools used to monitor
the performance of the LIGO interferometers and the data quality.
The centralized location of these automated analyses
also allows for detailed follow up of any identified issues. 

The raw HTML is
built programmatically through the \gwsumm software package~\cite{gwsumm}, an
extension of \gwpy that also manages core signal processing, while interactive
webpage elements are implemented through JavaScript. The visual layout of the
front end is color-coded by interferometer, with responsive web design
accomplished through Bootstrap~\cite{bootstrap} and a custom extension thereof
called GW-Bootstrap~\cite{gwbootstrap}.\footnote{To keep the file directory
structure clean, these packages are published through the Node.js Package
Manager (\texttt{npm})~\cite{npm} and supplied to the LIGO summary pages via
content delivery networks.}

While the LIGO detector characterization summary pages are built on-the-fly via Python code,
they are also designed to be easily tunable through configuration files. Users
have the freedom to register, design, and build a diverse array of
visualizations, ranging from simple timeseries tracks to more complicated
time-frequency spectrograms, with fine-grain control over all signal
processing parameters. Because the back end utilizes Asynchronous JavaScript
and XML~\cite{ajax} web development techniques, users also have the option to
build their own custom images and HTML on the server side, then load them
remotely through the summary pages. This workflow allows several third-party
analyses to be hosted in one centralized location, including automated
data-quality products for persistent gravitational wave searches.

While the \gwsumm software package is most heavily used by LIGO, it is
designed for use by the broader IGWN community. A suite of pages is currently
built using the same software for the 
KAGRA detector~\cite{Akutsu:2018axf,kagra_summary}, 
while independent software provides a very similar
service for the Virgo detector~\cite{virgo_summary}. A less extensive
public-facing version of the summary pages, built with \gwsumm and focusing
only on time segments and gravitational wave strain data, is also available
\cite{public_summary}.

\subsubsection{LigoDV-web}

The \ldvw service (LDVW)~\cite{ldvw} is an online data visualization
platform providing direct interactive access to data recorded at the LIGO
Hanford and Livingston observatories and a subset of data from
Virgo, KAGRA, 
the GEO600 observatory in Hanover, Germany~\cite{Grote:2010zz},
and the smaller 40m prototype
interferometer in Pasadena, CA, USA~\cite{Weinstein:2002cp}. 
This software instantaneously provides users with custom
visualizations of small data sets in a fast, secure, and reliable manner and
with minimal software, hardware, and training requirements. 
LDVW adds a convenient online tool that allows the generation 
and sharing of custom data visualizations to augment standardized 
analyses such as those on the Summary Pages. 
It is often the most convenient way to 
access the large number of different data sources at each site and
generate large numbers of plots to address specific questions.

LDVW is
implemented as a Java Enterprise application~\cite{java} with a proprietary
network protocol used for data access on the back end~\cite{nds}.
LIGO-Virgo-KAGRA Collaboration members with proper credentials can request data
to be displayed in several formats from any Internet appliance that supports a
modern browser with JavaScript and minimal HTML5 support, particularly personal
computers, smartphones, and tablets.
The primary signal processing and image rendering engine for this service is
\texttt{gwpy-plot}, a robust command-line interface for \gwpy~\cite{gwpy}.

\subsubsection{Data Quality Reports}
\label{ss:dqr}

In the context of detector characterization, a \dqr (DQR)~\cite{dqr} is an
internal collection of convenient analysis routines used to support and enable
the vetting of gravitational wave event candidates. It is tightly integrated with the \lvalert
(LVAlert)~\cite{LVAlert} and \gracedb (GraceDB)~\cite{GraceDB}. When an
upstream search pipeline identifies a potential gravitational wave signal, the event is
recorded in GraceDB and the LVAlert system broadcasts a notice to all
subscribers, including the DQR architecture. When the DQR receives an alert it
triggers a series of analyses from three LIGO computing clusters. 
Examples of included analyses are omega scans, statistical checks such as \hveto, 
and checks of known flags in DQSEGDB.  
The DQR infrastructure is modular, allowing for additional tools to be added
as desired.

Within minutes of the first automated notice sent to 
the \ac{GCN}~\cite{BarthelmyGCN} announcing the identification of a gravitational wave event candidate, 
the DQR architecture begins
to upload web-based reports and supporting data to GraceDB for internal review,
which then informs the decision to disseminate additional \ac{GCN} Notices and 
Circulars or to retract an announced candidate.
Additional details about the tools currently implemented in the \dqr
and related event validation procedures are described in \sref{s:validation}.

\subsubsection{Spectral artifact tools for persistent gravitational wave searches}
\label{sec:coh}
Several different tools have been developed to aid in finding narrow, 
persistent spectral artifacts in gravitational wave detector data~\cite{Covas:2018}.
These tools build amplitude spectral density plots using \acp{FFT} that are 
1800~s, or longer, over time periods of 1-day up to an entire observing run.
Since the coherent baseline is much longer than other figures-of-merit, and 
averaged across epochs, it allows for understanding the narrow, persistent 
spectral artifacts that corrupt searches for continuous gravitational waves from spinning 
neutron stars.

One of these tools, known as \texttt{Fscan}, runs automatically each day and
generates 1800-s-long \acp{FFT} for the low-latency gravitational wave strain channel and
a subset of auxiliary detector channels and physical environment monitoring
channels.
Various figures of merit can be derived from the \ac{FFT} data computed from 
the primary gravitational wave strain channel and the subset of additional channels.
This enables more regular monitoring of the behavior of spectral artifacts.

For example, normalized, day-, week-, and month-long averaged \acp{ASD} are
computed from the \ac{FFT} data as well as coherence between the gravitational wave strain
channel and the subset of additional channels.
Correlations of spectral artifacts in \acp{ASD} of different channels can then
be identified via these coherence measurements as possible non-astrophysical causes of
spectral artifacts.
Coherence is a useful figure of merit to reject spurious coincidence of
spectral artifacts that are not actually correlated and to identify potential
coupling mechanisms of non-astrophysical noise into gravitational wave data.

Specifically, coherence between two channels $d_1(t)$ and $d_2(t)$ is defined
as~\cite{Covas:2018}
\begin{equation}
\Gamma(f) = \frac{\langle | \tilde{d}^\star _1(f)\tilde{d}_2 (f)|^2\rangle }{\langle  | \tilde{d}_1(f) |^2  \rangle \langle  | \tilde{d}_2(f) |^2  \rangle  },
\end{equation}
where $\tilde{d}_i(f)$ ($i=1,2$) is the Fourier transform of the time series 
data $d_i(t)$, $^\star$ denotes complex conjugation, and the average $\langle 
\cdot \rangle$ refers to an average over $N$ segments.
For Gaussian, uncorrelated noise, the expected distribution of the coherence is 
given by
\begin{equation}
\label{eq:coherence-distribution}
p(\Gamma) \propto e^{-\Gamma N},
\end{equation}
where $N$ is the number of segments used for averaging.

Other examples of figures of merit include: \texttt{FineTooth}, a comb
identification and tracking tool; \texttt{NoEMi}, a line monitoring and
database tool; a coherence tool database that enables efficient look-up of
coherences in different time- and frequency-intervals; and studies that fold
time-domain data at periodic intervals to check for periodic elevated noise.
Additional figures of merit are under development for future use to aid
understanding of spectral artifacts in gravitational wave detector data.

\section{ Instrumental Investigations}
\label{s:noiseinvestigations}
On top of a simple Michelson interferometer, Advanced LIGO detectors employ a 
number of technologies to increase the overall sensitivity~\cite{aLIGO}. The 4 km long 
Fabry-Perot cavities, also known as X and Y arms, increases the total 
interaction time of the main laser beam with the gravitational wave. The use 
of recycling cavities enhances the detector's response to a passing 
gravitational wave. At the symmetric end of Michelson, the power recycling cavity amplifies 
the total laser power transmitted to the LIGO arms. The signal recycling cavity on the 
anti-symmetric end is used to vary the frequency response of the detector. 
Figure~1 in \cite{Buikema:2020dlj} contains a detailed schematic of the 
Advanced LIGO detector.
Additional important subsystems for the operation and 
characterization of the LIGO detectors are the phyiscal environment
monitoring subsystem, 
which includes a large array of sensors that track any changes to the environment
around the detectors~\cite{Effler:2014zpa,Nguyen:2021ybi},
and the squeezer subsystem, 
which is used to inject squeezed light into the detectors and reduce
residual noise from the quantum state of 
the laser~\cite{Snabel:2010sq,Barsotti_2018,Tse:2019wcy}.

In order to maximize the opportunities for the discovery of astrophysical 
gravitational-wave signals, it is essential to understand the instrumental and 
environmental noise that can mimic or obscure such signals. Recognition of 
potential noise couplings can lead to detector hardware changes to reduce the 
rate of noise artifacts. In this section, we first describe our approach to 
identifying instrumental noise and mitigating its 
effects on astrophysical searches. Later we discuss the major types and sources 
of transient noise and their impact on detector data quality. 


\subsection{Instrumental Investigation methods}
\subsubsection{Data Quality Monitoring}

The LIGO instruments are delicate in the sense that glitches or other 
manifestations of noise can appear in the gravitational wave strain channel. 
During the observing period,  a ``shifter'' assigned at each site 
conducts a week long data quality shift.  The objective of the 
data quality shifter is to monitor the behavior of the instruments, 
note any changes, and to communicate them to the commissioners at 
\ac{LHO} and \ac{LLO} and the members of the detector characterization group. 
The LIGO summary pages~\cite{ligo_summary} are 
the typical
launching point for off-site data quality investigations of instrument noise. These pages 
provide both an overview of the detector status and the low-level information 
about specific detector subsystems. The summary pages also display the results 
of analysis algorithms that identify or correlate noise in both auxiliary 
sensors and gravitational-wave strain data.  The computing infrastructure of 
the summary pages is discussed in further detail in \sref{summary_pages}.

Omega scans, \hveto, Lasso \cite{Walker:2018ylg,gwdetchar} and Omicron are some of 
the most commonly used analysis tools for identifying noise in the detector.  
Tracking down sources of transient noise often begins with the output of
Omicron (see \sref{omicron}), which finds short-duration bursts of 
noise.  The resulting events, or ``triggers'', can be plotted in the 
time-frequency plane to visualize
transient noise in both witness auxiliary sensor and gravitational wave strain data. For each day, 
the Omicron triggers
(glitches) for the gravitational wave channel are shown on a time-frequency plot with the 
markers
color-coded for \ac{SNR}. An increase in the number of high \ac{SNR} glitches 
indicates a
change in the instrument or environment. If glitches persist at specific 
frequencies critical
to data analysis, they constitute a problem for event detection and parameter
estimation. Therefore, it is essential to eliminate the harmful influence of 
these glitches on the
searches for events. To recognize the potential sources of problematic 
glitches, we use 
\hveto (see
\sref{hveto}) to identify witness auxiliary sensors whose bad behavior coincides with the 
appearance of a subclass
of glitches.  
\begin{figure}[t]
    \centering
    \includegraphics{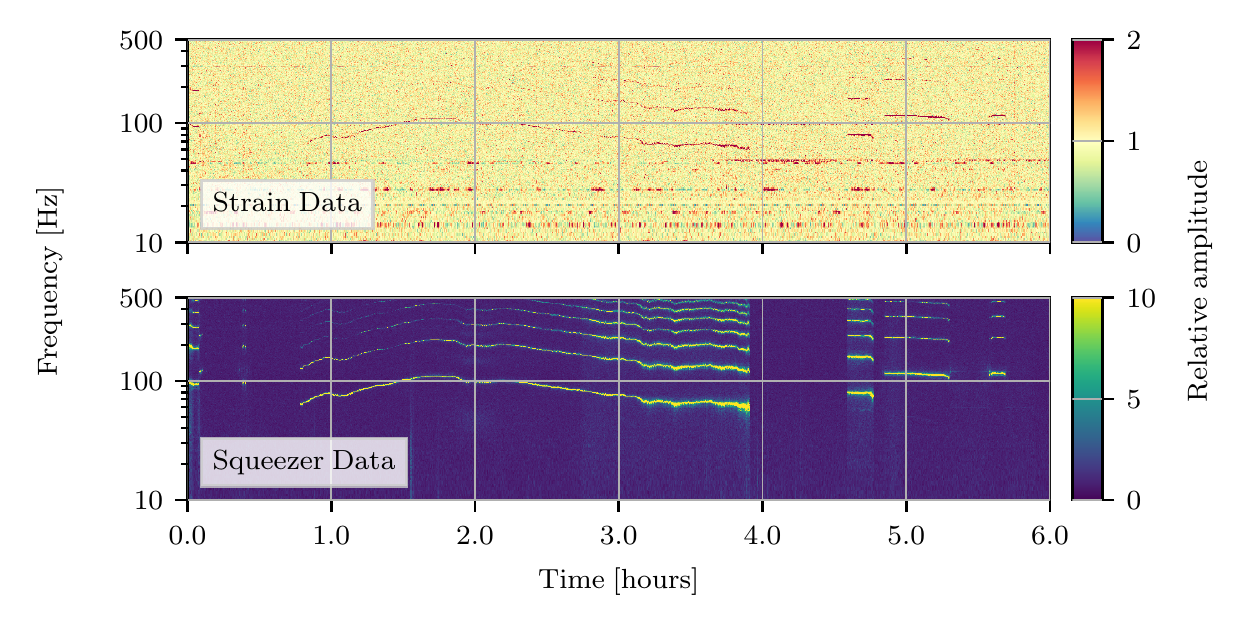}
    \caption{\textit{Top}: Spectrogram of the relative amplitude of the gravitational-wave strain 
during the  first 6 hours (UTC) of 2019-04-26. Between about 3h and 6h, at 
frequencies of about 40 Hz to 300 Hz, peculiar line features are visible. These 
are the wandering (in frequency) lines discussed in the text. \textit{Bottom}: 
Squeezer data when the wandering line in the gravitational-wave strain is visible. Features (in 
yellow) in this squeezer channel match those seen in the gravitational-wave strain in time and 
frequency. See~\cite{alog:sqz_noise_eater} for details.}
    \label{fig:squeezer}
\end{figure}

As an example of the value of data quality monitoring, we consider the ``squeezer 
wandering line'' at \ac{LHO} as described here 
~\cite{alog:sqz_noise_ht,alog:sqz_wandering_line}. A line-like feature with 
changing frequency was clearly visible in hourly spectrograms of the gravitational-wave strain (see 
\fref{fig:squeezer}). This equally spaced comb of lines will appear and 
disappear suddenly between 80 Hz and 140 Hz at \ac{LHO}. Similar wandering lines 
were noticed at LLO between 160 Hz and 200 Hz. This feature was shown to be 
correlated in time and frequency with a strong feature in several squeezer 
channels at both the 
detectors. The data quality issue was solved by turning off the amplitude stabilizer 
for the laser used to provide squeezed light at \ac{LLO}~\cite{alog:sqz_noise_eater}. 
This was then also implemented 
successfully at \ac{LHO} to cure the 
problem ~\cite{alog:sqz_noise_ht_jenne}. 
See~\cite{Berger:2018ckp,Nuttall:2018dq} for examples 
of additional solved transient noise issues.

\subsubsection{Physical Environment Monitoring}
\label{ss:pem}
Environmental noise can affect a LIGO detector by limiting its sensitivity to 
astrophysical gravitational wave signals and producing transients in the strain data. Some 
environmental noise sources can potentially be correlated between detector 
sites, making them particularly problematic for astrophysical searches. It is, 
therefore, important to identify these sources and mitigate their effects. The 
methodology and hardware for investigating environmental noise are discussed in 
detail in ~\cite{Effler:2014zpa,Nguyen:2021ybi}. They typically consist of generating a 
``noise injection'' of known amplitude and frequency range and observing the 
detector's response to the signal. Sensors that monitor the physical 
environment are used to measure the injection and estimate the noise source's 
coupling to the detector. For example, we estimate acoustic coupling by using 
accelerometers and microphones to measure acoustic noise injections made from 
speakers. The sensors are much more sensitive to environmental noise than the 
detector is, making them good witnesses of noise sources that couple into the 
gravitational wave strain channel. Other injection methods include shaking injections for 
studying seismic coupling, magnetic field injections for coupling due to 
permanent magnets and electronics, and radio-frequency electromagnetic 
injections for radio frequencies
coupling to electronics in the interferometer controls. Coupling functions are 
used to assess the need for mitigation as well as to estimate the contribution 
of transients to the gravitational wave strain channel, such as when validating gravitational wave 
events 
~\cite{Schofield:GW150914,Schofield:GW170817}.

Environmental investigations were used to track down the source of a $48$~Hz 
peak in the gravitational wave strain channel at LHO throughout 
\ac{O3a}~\cite{alog:pem}. 
Injections using shakers and speakers showed that the noise was originating 
from somewhere in the corner station. Further investigation using physical 
impulses pointed specifically to the area near the vertex area. A new 
technique, whereby two shakers inject sine waves at slightly different 
frequencies to produce beats in the injection amplitude, showed that the beats 
in the motion of a particular vacuum chamber door correlated the most with the 
resulting beats in the gravitational wave strain channel. The noise was found to be the 
result of 
scattered light through the chamber viewports and was promptly mitigated by 
blocking the scattered beams, eliminating the $48$~Hz peak from the gravitational wave 
channel~\cite{Nguyen:2021ybi}.

\subsubsection{Safety studies}\label{s:safety}
Many LIGO detector characterization analyses aim to ensure that 
gravitational-wave candidates are astrophysical and not caused by terrestrial 
noise. These analyses, such as the iDQ framework described in
\sref{ss:low_lat_dq}, typically search for statistical correlations between
auxiliary channels measuring the environment surrounding the detectors and 
the gravitational wave strain channel. If strong 
correlations are found, these gravitational wave candidates may be attributable to 
environmental noise and hence vetoed. This process breaks down, however, if 
there are auxiliary channels which pick up disturbances in the gravitational wave channel. In 
that case, an astrophysical signal may appear in both the gravitational wave channel and such an 
auxiliary channel, and the signal may subsequently be erroneously vetoed due to 
the channel's correlation. Hence, information about the coupling of auxiliary 
channels to the gravitational wave channel is essential. Auxiliary channels which may record 
excess power originating in the gravitational wave channel are considered ``unsafe'' for vetoes, 
and the channels which are not found to be coupled in this way are then 
classified as ``safe.'' Only ``safe'' channels are used in the vetoing of gravitational wave
candidates. These categorizations of channels are referred to as``safety'' studies.

Since transfer functions between the strain channel and most auxiliary channels 
are not well known or understood, channel safety is determined empirically via 
hardware injection safety studies.
These are conducted by injecting sine-Gaussian signals of various frequencies 
and amplitudes directly into the strain channel. 
In \ac{O3}, a new set of injections was designed with wider time intervals 
between 
injections, since event trigger generators such as Omicron and KleineWelle~\cite{kleinewelle},
used in the subsequent safety analyses of the hardware injections, were found 
to be unable to distinguish between successive injections when they were spaced
fewer than about three seconds apart.
These injected signals range in frequency from 20~Hz to 700~Hz, with amplitudes
corresponding to \acp{SNR} ranging from 15 to 500~\cite{LLO_inj_alog, LHO_inj_alog}.
Each signal of a certain frequency and amplitude was injected three times,
spaced five seconds apart.

Algorithms such as the \pointy statistic~\cite{Essick:2020cyv}
and \hveto then run analyses on the hardware injections to generate lists of 
safe and unsafe channels.
\pointy is a null-test that uses the assumption that events in auxiliary 
channels are distributed according
to stationary Poisson processes. 
For each channel, the Poisson rate is measured using a time window much larger 
than that of the 
hardware injections.
Then, using the measured rates and a set of significance thresholds, \pointy 
produces a p-value timeseries
for each auxiliary channel sampled above 16~Hz.
Auxiliary channels that have anomalously small p-values
\footnote{This is typically chosen to be about \DANGERPVALUE.}
at the injection times 
are then declared to be unsafe,
and the other channels are declared safe.

\hveto~\cite{Smith:2011an,hveto} (see \sref{hveto}) correlation 
searches are used to compare all injections to each channel analyzed by 
Omicron~\cite{Robinet:2020lbf} in daily operation. 
The operation of the safety-oriented \hveto search is the same as described in
\sref{hveto} without the hierarchical removal of time periods
with identified correlations.
This difference allows all statistically significant correlations with the injection
set to be identified, even if the auxiliary channel data streams are themselves
highly correlated.  
Channels with high significance are then 
visually inspected using an
omega scan and 
glitchgrams~\cite{gwpy} to distinguish between witnessing the injection from 
chance. 
A low threshold for significance to trigger manual follow-up is used to 
minimize the risk of identifying an unsafe channel as safe.

Channel safety lists are then compiled using the results of the \pointy and 
\hveto studies.
In \ac{O3}, these two tools largely agreed on safety results, with most 
differences 
arising from the larger pool of channels
analyzed by \pointy. 
When disagreements were identified which could not be reconciled based on 
expected false alarm rates, we erred on the side of caution, declaring channels 
unsafe even if only one algorithm classified it as such.
At each detector, approximately \UNSAFECHANNELSPERDETECTOR channels were declared unsafe,
representing about \PERCENTOFCHANNELSUNSAFE~\% of the 
$\mathcal{O}$(\TOTALCHANNELSANALYZED) auxiliary channels
sampled at or above $16~\mathrm{Hz}$ that were analyzed.

Beyond channels which are expected to be unsafe, such as channels in the 
\ac{DARM} control loop and many of the suspension channels, the safety
algorithms also identified a number of channels which we would not have
expected \emph{a priori} to be unsafe. 
For example, a set of the channels in the alignment sensing and control subsystem at \ac{LLO}
measuring an radio frequency photodiode at the
anti-symmetric port were found to be unsafe.
These channels do not directly measure the gravitational wave strain, and were thought to 
be far
enough away from the sensing channels to be safe, but particularly loud
injections were able to excite them, resulting in an unsafe classification.
Even the loudest injections possible in these studies 
($\mathcal{O}(500)$ \ac{SNR}), however, are not necessarily loud enough to 
excite all potentially unsafe channels.
These include channels such as magnetometers in the electronics bay suspension 
rack, which are all suspected to be unsafe at high enough SNRs.
The magnetometers are not themselves coupled to gravitational wave strain, but are near 
electronics which drive the actuation to the test mass mirrors.
A sufficiently loud signal in the interferometer would require significant
actuation from these electronics to keep the test mass mirrors still, which 
would then be detected by the magnetometers. 
This phenomenon is also observed in the channels which monitor the electrostatic drive 
power supply, as detailed in~\cite{alog:vol_mon}.
We do not foresee these potential oversights in safety classification due to 
the limitations of hardware injections becoming a problem in gravitational wave veto
analysis, since gravitational wave signals louder than the hardware injection limit are 
not expected.

\subsection{Known classes of instrumental noise}

\begin{figure}[t]
    \centering
    \includegraphics[width=\textwidth]{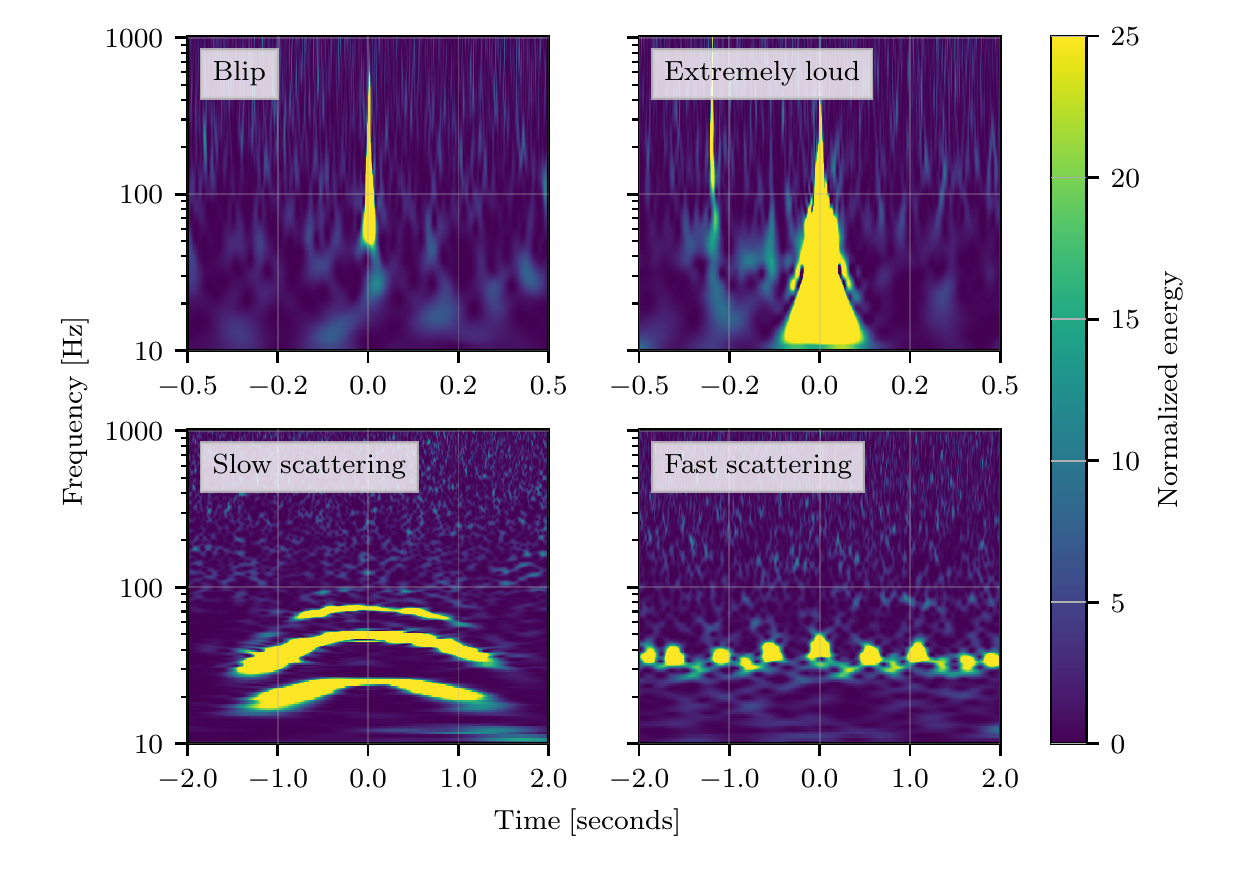}
     \caption{\textit{Top left}: Blips are short duration transients with large 
bandwidth and can mimic a gravitational-wave signal of high mass compact 
binaries. \textit{Top right}: An example of a loud trigger, classified as 
Extremely$\_$Loud by the GravitySpy. These high SNR triggers often cause large 
drops in the astrophysical range and adversely affect the sensitivity of the 
detector. \textit{Bottom left}: Slow 
scattering caused by ground motion in the microseism band ($0.1-0.3$~Hz). 
Multiple reflections between the test mass and moving surface can generate 
higher frequency harmonics as seen here. \textit{Bottom right}: 
Fast scattering triggers caused by ground 
motion in $1 - 6$~Hz band. Trains, human activity, and logging near 
the site are 
the most common causes of fast scattering noise.}
     \label{fig:glitch_classes}
\end{figure}

Despite the wealth of information available from the LIGO summary pages and 
automated algorithms to correlate noise with auxiliary channels, several 
classes of glitches have persisted in the data with insufficient clues to 
remove them all. Here, we describe some of the most frequently occurring 
transient noise and our efforts to identify/mitigate the noise coupling. The 
three most common types of glitches are Blips, Light scattering, 
and Loud triggers~\cite{LIGOScientific:2018mvr,TheLIGOScientific:2017lwt,
Zevin:2016qwy}. Spectrograms of each of these glitch classes produced using
the Q-transform are shown in 
\fref{fig:glitch_classes}. These glitches are typically categorized based 
on their time-frequency evolution. This task is achieved for a large number of 
glitches through GravitySpy~\cite{Zevin:2016qwy}, 
a machine learning framework that uses the 
convolutional neural networks to classify transient noise based on the glitch 
morphology in the spectrogram. For the GravitySpy project, members of the detector 
characterization group helped create an initial dataset by identifying major 
glitch categories. Currently, this algorithm classifies transient noise into 23 
different classes.  
The data set used to train GravitySpy, which includes information about glitches
in publicly available LIGO gravitational wave strain data, 
is available~\cite{coughlin_scott_2018_1476551}. Certain classes of transient
noise, such as Blip, Tomte, Extremely Loud, and Koi-Fish, have not shown any
environmental 
or instrumental coupling yet.  Statistical data analysis of these glitch 
categories has led to an improved noise characterization, and we continue to 
look for the 
source of the noise. On the other hand, Scattered Light noise has shown a 
strong environmental coupling through the ground motion near the detectors.
For one of the populations of noise due to light scattering, we were able to 
find the exact noise coupling through instrument investigations. Instrument 
changes implemented during O3b fixed the noise source, which led to an 
improved detector performance during high ground motion. We discuss this in 
more detail in \sref{ss:scatter}.

\subsubsection{Blips}
\label{ss:blips}
Blip glitches are subsecond duration glitches with high frequency bandwidth and 
no known instrumental or environmental coupling. Due to its appearance in 
time-frequency plane, as shown in the top-left of \fref{fig:glitch_classes}, 
a blip glitch may resemble a 
gravitational wave signal from the merger of high mass binaries.
Blips occurred with a rate of approximately \BLIPRATE at \ac{LHO} and \ac{LLO} 
during the 
second observing run. This rate increased to about \BLIPRATELLO at \ac{LLO} 
during 
\ac{O3}, while no significant changes were observed for LHO. Blip glitches are 
responsible for a significant portion of the unvetoed high SNR 
background, reducing the effectiveness of both modeled and unmodeled searches 
for gravitational waves~\cite{Nitz:2017lco}. These short duration transients 
appear to have multiple subcategories that may be caused by different physical 
mechanisms, but distinguishing different types can be difficult given their 
short duration and simple signal morphology. In the GravitySpy citizen science 
project, there are other classes of glitches (called “Koi fish” and “Tomtes”) 
that have similarities to blips and may be related. Many 
investigations 
have been undertaken to find clues for their origin, but so far no conclusive 
evidence has been found to explain a mechanism for all of them 
\cite{alog:blipH1,alog:blip_statistics}. Weather conditions at \ac{LHO} 
during \ac{O3} prevented the correlation between low humidity and high blip 
glitch rate from being observed as suggested in 
\cite{Cabero:2019orq,dcc:blip_LHO_Adrian}. Low energy cosmic rays striking the 
LIGO mirrors are responsible for a minute amount of noise in the 
interferometers~\cite{Yamamoto:2008fs,Braginsky:2005ai}. High energy cosmic rays 
could cause blip glitches by strongly perturbing the detector's mirrors. By 
studying temporal differences between cosmic ray strikes and blip glitches at 
\ac{LHO}, it was found that cosmic rays were not correlated with blip glitches 
in \ac{O2} or \ac{O3} in \ac{LHO}~\cite{dcc:blip_LHO_Adrian}.  

After the conclusion of the 
third observing run, the light source in \ac{LHO} was shut off to determine 
whether 
blip glitches occurred as a result of errors or data corruption in the 
interferometer's data acquisition system~\cite{alog:IMC_shutdown}. No blip 
glitches were observed in two output mode cleaner photodiode channels, which regularly 
feature 
blip glitches during routine operations over this time period, 
indicating that data acquisition processes that depend on the output mode cleaner photodiodes are 
unlikely to be the 
source of blip glitches. However, some subsets of blips have been found to be 
correlated with instrumental issues, including computer timing errors, but this 
only explains some percent of them~\cite{Cabero:2019orq,alog:blipH1DCPD}.

\subsubsection{Loud Triggers}\label{sec:loudtriggers}
Loud triggers in LIGO refer to short duration transients with very high 
\ac{SNR} (\ac{SNR} $>$ 100). These loud noise transients are associated with 
massive drops in the detector's astrophysical range, adversely affecting its 
sensitivity.
GravitySpy classifies most of these triggers as Extremely Loud, followed by Koi 
Fish.  As of yet, we have not found a coupling for these loud glitches in the 
detector and they are consistent with a Poisson distribution.  During O3, they 
occurred with an average rate of $3.3$ per hour at \ac{LLO} and $3.5$ per hour 
at LHO, with a minor reduction in the rate observed in O3b compared to O3a.  As 
the top-right of \fref{fig:glitch_classes} shows, 
a loud glitch often saturates the time-frequency 
spectrogram in the band $10-500$~Hz.  These triggers are witnessed by some 
length sensing and control channels very frequently, and the \hveto tool 
on the summary page consistently finds statistical correlations between some 
length sensing and control channels and loud triggers in gravitational wave strain. We are 
investigating if any issues in the length sensing and control feedback loop have any causal 
association with these very loud triggers in the gravitational wave strain channel. 
Fluctuations in 
voltage monitors coincident with loud triggers 
in the gravitational wave strain channel have been ruled out as a 
possible cause~\cite{alog:vol_mon}.

\subsubsection{Scattered Light}\label{ss:scatter}

Another class of noise that frequently appears in both detectors is 
caused when a small fraction of laser light gets scattered off of the 
test mass, hits a moving 
surface, also known as scatterer, and then rejoins the main beam. The time 
dependent phase difference between the main beam and the scattered beam, 
caused by the relative 
motion between the test mass and the scatterer, introduces noise in the 
gravitational wave channel. This noise due to light scattering shows up as 
arches in the time-frequency spectrograms. Its amplitude depends on the 
amount of scattered light that recombines with the main beam, while the maximum 
frequency of gravitational wave strain impacted by arches is a 
function of the strength of relative motion between 
the two surfaces. During O3, we observed two different populations of 
scattering transient noise: Slow Scattering and Fast Scattering.

Slow scattering triggers refer to the long duration arches visible in the 
time-frequency spectrograms during high ground motion in the microseism band 
($0.1-0.3$~Hz). Depending upon the amount of ground motion, these triggers 
would affect gravitational wave strain sensitivity in the band $20 - 120$~Hz. 
Moreover, during periods of particularly intense microseismic activity, higher 
frequency harmonics of the scattering arches can be seen in the spectrogram 
indicating multiple reflections of the scattered light beam between the test 
mass and the scatterer. This can be seen in the bottom-left of
\fref{fig:glitch_classes}.
During O3, we found two distinct paths in the detector through which this noise 
would couple to the primary gravitational wave strain channel. The identification and 
mitigation of 
these noise couplings are discussed later in this section. 

Fast scattering triggers, as shown in the bottom-right of \fref{fig:glitch_classes}, 
are strongly 
correlated with ground motion activity in 1 to 6~Hz band. Human activity near 
the site, trains near the Y end of \ac{LLO}, and thunderstorms near the site are 
known 
to increase the rate of fast scatter \cite{dcc:fast_sc_sidd}. As compared to slow scattering, these 
triggers typically have higher peak frequency, lower SNR, and lower duration. 
Due to differences in gravitational wave strain sensitivity and ground motion in $1 - 
6$~Hz band, 
this particular class of noise is a lot more frequent at \ac{LLO} than at 
\ac{LHO}. 
 
The LIGO end stations located at the end of 4 km long arms contain the seismic 
isolation system, \ac{TMS} to monitor the transmitted light, and seismometers 
to measure ground motion. The seismic isolation system includes a quadruple 
main and reaction pendulum also known as the main and reaction chain 
respectively. The bottommost mirror on the main chain and the reaction chain 
are known as \ac{ETM} and \ac{AERM}.  To keep the interferometer on resonance, 
\acp{OSEM} and electrostatic drives installed on the reaction chain masses are used for 
applying control signals on the main chain \cite{Aston_2012}. A small amount of 
light transmitted through the \ac{ETM} is received by the photodiodes on the 
\ac{TMS}, located behind the suspension assembly at each end station. These 
photodiodes are used for monitoring the transmitted field that contains the 
information of the field circulating in the arm cavities. Figure 2 in 
\cite{Soni:2020rbu} contains a schematic of LIGO end station housing.

A collection of \acp{OSEM} are 
positioned throughout the LIGO interferometers, capturing the motion of several 
optical components likely to scatter laser light. The 
\texttt{gwdetchar-scattering} algorithm~\cite{gwdetchar,Accadia:2010zzb} 
identifies time segments in which the motion of these \acp{OSEM} can be 
projected between $10-60$~Hz, then produces a webpage displaying significant 
gravitational wave strain 
Omicron triggers in this frequency band during these time segments. This tool 
also generates an omega scan for a random sample of such Omicron triggers. 
Information for each UTC day of the observing run is stored as part of the LIGO 
summary pages, enabling analysts to not only identify scattering triggers 
but also to trace the broader origin of noise due to scattered light.

\begin{figure}[t]
\centering
\includegraphics[width=\textwidth]{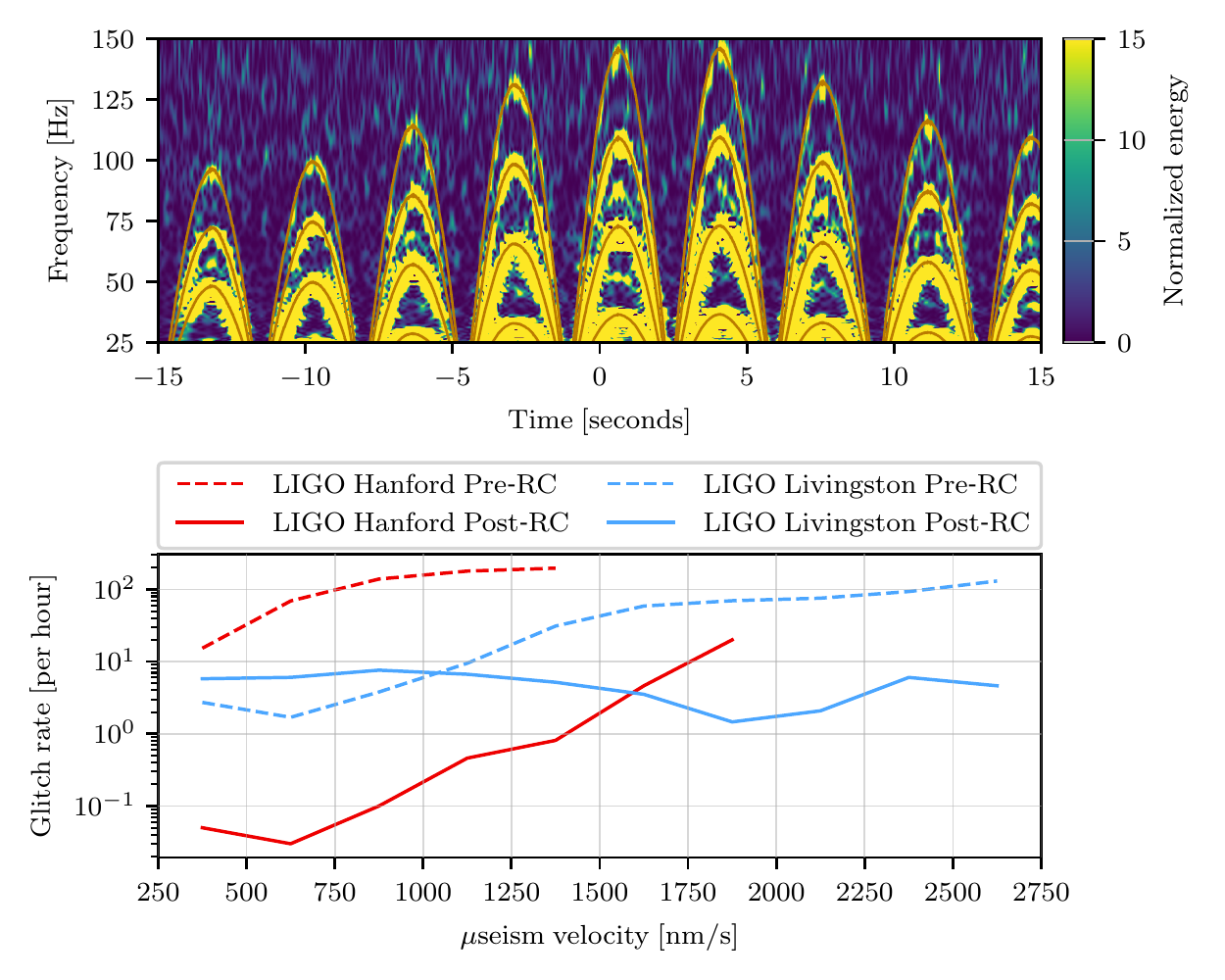}
     \caption{\textit{Top}: Fringe frequency motion as recorded by 
the penultimate (L2) stage 
OSEM of quad suspension overlaid on the slow scattering arches in gravitational wave 
strain. This 
OSEM measures the distance fluctuation between the main chain and the reaction 
chain of the suspension at the L2 stage. \textit{Bottom}: A 
comparison of the rate of slow scattering triggers for pre and post Reaction 
chain (RC) tracking at \ac{LHO} and \ac{LLO}. RC tracking resulted in a 
significant 
drop in the rate of scattering at both the sites. At LHO, 
ground motion did not exceed 1500 $\mathrm{nm}$/s for Pre RC and did not exceed 
2000 $\mathrm{nm}$/s for Post RC tracking.}
     \label{fig:fringe_glitch}
\end{figure}

The scattering summary page hinted at a correlation between ground motion 
in $0.03 - 0.3$~Hz band and motion in the \ac{L2} stage \ac{OSEM} of the quad 
suspension. As shown in the top plot of \fref{fig:fringe_glitch}, we also noticed a strong match 
between the fringe 
frequency motion of the \ac{L2} stage and the slow scattering arches in gravitational wave 
strain spectrogram. 
Other than this, \hveto suggested a statistical correlation between slow 
scattering in gravitational wave strain and noise in the transmitted light monitors on the 
\ac{TMS}. A follow-up data quality investigation 
confirmed the existence of these two noise couplings, both via the \ac{ETM}. 
The first coupling, which came to be known as 
\ac{ETM}-\ac{AERM} scattering is due to the scattered light between the 
\ac{ETM} and \ac{AERM}.  A technique called 
\ac{RC} tracking, implemented in Jan 2020 at \ac{LLO} and \ac{LHO}, reduced 
the distance fluctuations between the \ac{ETM} and \ac{AERM}, which was found 
to be the 
source of noise in gravitational wave strain. Following this RC tracking at both the 
detectors, the slow scattering glitch rate was reduced for ground motion above 
1000 $\mathrm{nm}$/s in the microseismic band. This is shown in the bottom plot 
of \fref{fig:fringe_glitch}. The second, comparatively weaker noise coupling 
is due to scattered light between the \ac{ETM} and \ac{TMS}, and is known as 
\ac{ETM}-\ac{TMS} scattering. \ac{TMS} tracking, to reduce the 
relative motion between the \ac{ETM} and \ac{TMS} is in place and will be 
activated before the next Observing run~\cite{Soni:2020rbu}.

\begin{figure}[t]
    \centering
    \includegraphics[width=\textwidth]{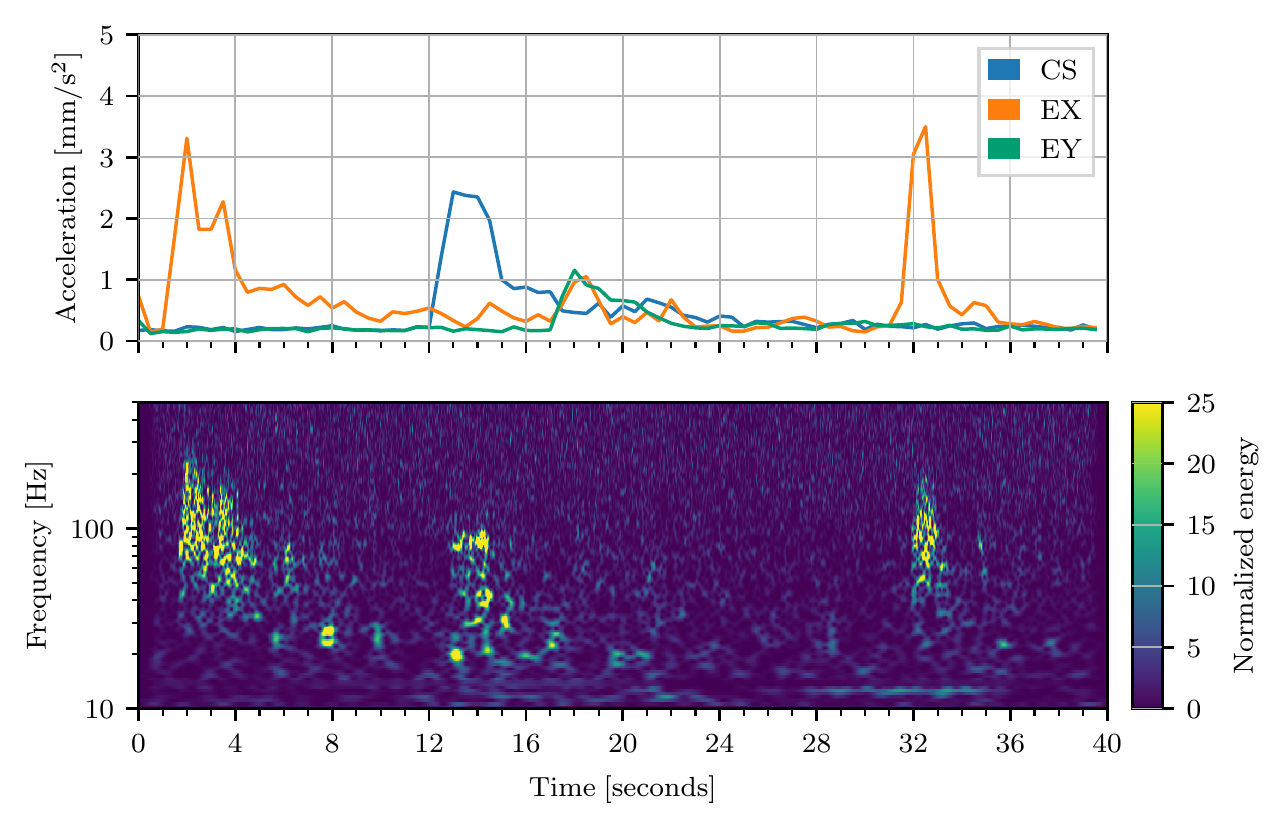}
    \caption{\textit{Top}:\ac{RMS} value of the 
accelerometer signals after applying a band-pass filter with a frequency range 
10~Hz to 100~Hz, where thunder manifests as peaks. 
\textit{Bottom}: Spectrogram of the gravitational wave strain channel at the time of the 
same 
thunderstorm. Excess 
noise in the frequency range of 20~Hz to 200~Hz coincides with the 
thunderclaps, with intensity depending on the thunder's location.}
    \label{fig:thunderstorms}
\end{figure}

\subsubsection{Thunderstorms}
\label{thunderstorms}
Thunderstorms are a common meteorological phenomenon in the south of Louisiana, 
where
\ac{LLO} is located. Depending on the distance from and nature of the 
lightning strikes,
the thunderclap can range from a sharp, loud crack to a long, low rumble. The 
thunderclap is
registered by the microphones, while the rumble is recorded by the ground 
motion sensors, such
as accelerometers. The thunder shakes the vacuum chambers enclosing the mirrors 
and laser
beam. One mechanism through which thunderstorms couple into the 
gravitational-wave
detector happens when scattered light is reflected from the chamber walls and 
recombines to
the main laser beam, producing excess noise in the gravitational-wave channel 
proportional to the
rumbling intensity. A thunder-driven vibration was used to show that the 
coupling estimates from physical environment monitoring injections (see \sref{ss:pem}) correctly
predict a strong coupling between the acoustic noise at the Y-end and gravitational wave strain 
noise at \ac{LLO} \cite{alog:coupling_EY}.

In the presence of thunder, the microphones and accelerometers attached to the 
vacuum
chambers register disturbances at frequencies below 200~Hz, with more intensity 
in the band 10~Hz to 100~Hz. 
These acoustic and ground motion perturbations coincide with 
noise in the
gravitational-wave channel at frequencies between 20~Hz to 200~Hz. The excess 
noise produces
drops in the sensitivity, which is quantified by the BNS range. During the 
third Observing run,
thunderstorms near the \ac{LLO} detector provoked a range decrease of 
approximately
15\%~\cite{thunder_LIGO}.

Thunder is unpredictable and, depending on the location of lightning, couples into 
the gravitational wave strain channel
through one or multiple areas simultaneously. The delay in arrival time of the 
rumble across the
different buildings allows for the localization of the source. This could be 
used for posterior studies about coupling mechanisms, although challenging 
to achieve due to possible multiple claps of
thunder happening at the same time~\cite{acoustic_LIGO}.
The plot at the top in \fref{fig:thunderstorms}, shows the root-mean-square 
value of the accelerometer signals after applying a
band-pass filter with a frequency range 10~Hz to 100~Hz, where thunder manifest 
as peaks in these
data. The bottom plot shows the spectrogram of the gravitational wave strain channel at 
the time 
of the same
thunderstorm, where the excess noise is coincident with the thunderclaps.

\subsubsection{Optical lever loud glitches}
Using the summary pages to monitor the instruments helped solve several 
instrumental 
problems during \ac{O2}. One example was a series of loud glitches in the 
gravitational-wave channel in \ac{LHO} caused by glitches in the power of 
lasers 
used as \acp{oplev}. 
Optical levers consist of auxiliary lasers aiming light at a 
mirror 
to allow alignment at the nanoradian level~\cite{TheLIGOScientific:2014jea}. 
If the light contributes enough photon 
pressure 
to the mirrors in an unsteady or glitchy way, this disturbance can appear in 
the gravitational wave 
channel~\cite{alog:H1oplev5}. 
It was first noticed in November 2016 
that glitches in the \ac{ETMX} \ac{oplev} were coupling to 
the gravitational wave strain~\cite{alog:H1oplev1}. 
While 
adjustments could be made to the laser power to reduce the glitching eventually, 
the laser was replaced in June 2017~\cite{alog:H1oplev2}. However, 
glitches continued to appear in the gravitational wave strain from this cause. 
\hveto found statistical correlations 
between these glitches in gravitational wave strain and noise in the Y-end optical lever. 
A veto definer flag (see section~\ref{s:flags}) flagged 
gravitational wave strain glitches with \ac{SNR} $>$ 65 and in the 10~Hz - 50~Hz
band coincident with the noise in ETMY oplev 
channel. Note that these glitches appeared even though the optical lever signal 
was not being used in the feedback loops controlling the mirror positions. The 
glitches only disappeared when the ETMY oplev was turned 
off~\cite{alog:H1oplev3}.

\subsubsection{Whistles}
Glitches caused by radio frequency beat notes also referred to as ``whistles'', are a 
common 
source of instrumental noise coupling into the gravitational wave strain 
channel.  They frequently appeared at \ac{LLO} 
during \ac{O2}, and in both \ac{LLO} and \ac{LHO} 
during \ac{O3}.  Whistles are typically identifiable by their characteristic 
``V'' 
or ``W'' shape in frequency vs. time spectrograms. However, their frequency 
content and timescale vary greatly over time, meaning some very short duration 
whistles are difficult to resolve based on their morphology.  Burst transient 
searches have especially been affected by whistle glitches in the last two 
observing runs, though matched filters can also be fooled by the part of the 
whistle, which increases in frequency over time in a manner similar to a 
compact binary merger.  Outbreaks of whistle glitches were observed to 
correlate with oscillations in the pre-stabilized laser frequency stabilization servo loop, 
usually correlated with seismic activity.  The issue was eventually mitigated 
by implementing a 
script to automatically adjust the pre-stabilized laser frequency stabilization system 
common gain~\cite{alog:whistle}.  
Fortunately, whistles typically couple into various auxiliary channels 
simultaneously with the gravitational wave strain channel, so for whistle 
glitches in existing data it is generally relatively straightforward to 
construct data quality flags. Omega scans of times associated with whistles in the 
gravitational-wave strain channel show coincident whistles in several channels 
in the ``length sensing and control'' and ``alignment sensing and control'' subsystems.  

\subsubsection{\label{sec:schumann}Schumann resonances and other magnetic coupling}
An important potential source of environmental correlations between gravitational wave
detectors are global magnetic fields including Schumann resonances~\cite{SchumannRes}.
Schumann resonances are modes in the effective resonant cavity formed by the Earth's surface
and ionosphere and are excited by lightning strikes. 
These roughly broadband magnetic fields can couple to the gravitational wave strain channel
via a variety of mechanisms, such as coupling to permanent magnets mounted on optics~\cite{Thrane:2014yza}.
We have established in \ac{O3} that the magnetic noise budget is well below the sensitivity of 
\ac{GWB} searches~\cite{dcc:isotropic_andrew}.
The only regular coupling to the gravitational wave strain channel from individual lighting strikes
observed so far is vibrational coupling to thunderclaps (see section~\ref{thunderstorms})
In the future, however, it will become increasingly important to develop a
detailed understanding of magnetic correlations in order to be able to make
confident statements about astrophysical sources of correlation.

A correlated magnetic signal can lead to an effective, non-astrophysical \ac{GWB}
signal, potentially mimicking an astrophysical \ac{GWB} signal.
To study this possibility, we compute the coherence between magnetometers located at two
sites, denoted as $M_{12}(f)$.
Additionally, we perform measurements of the coupling between magnetometer signals and the 
gravitational wave strain channel.
These measurements allow us to compute a magnetic noise budget
that can be compared with the stochastic search sensitivity, as described 
in~\cite{Thrane:2013npa,Thrane:2014yza,Coughlin:2016vor,Cirone:2018guh}.

\begin{figure}[t]
  \centering
  \includegraphics[width=\textwidth]{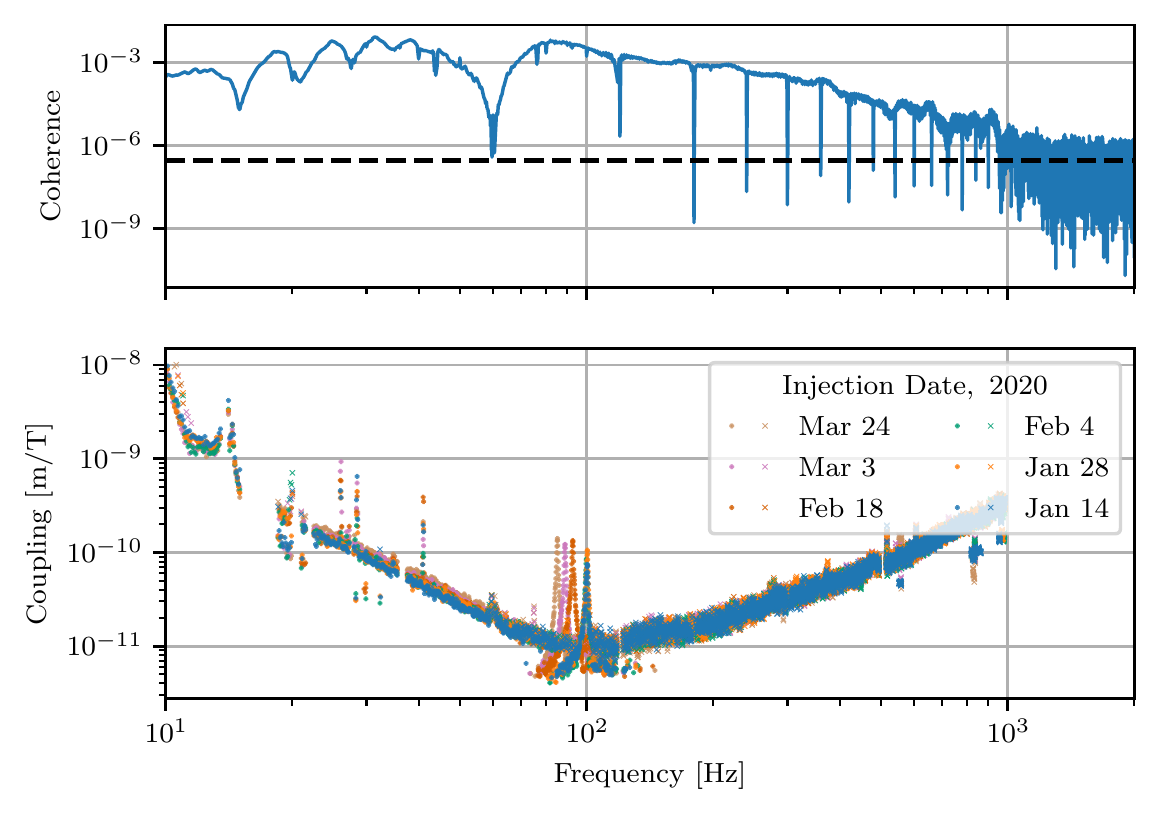}
  \caption{A figure showing magnetic correlations and coupling contributing
    to a correlated magnetic signal.
    \textit{Top:} The coherence, $M_{12}(f)$, between magnetometers at
    \ac{LHO} and \ac{LLO}.
    The peaks in coherence below 100 Hz are due to Schumann
    resonances.
    The black dashed line indicates the expected value of the coherence for
    uncorrelated, Gaussian noise.
    \textit{Bottom:} The magnetic coupling function, $C(f)$, at
    \ac{LHO} in units of meters per tesla. Measured values of $C(f)$ are denoted by dots
    and upper limit estimates on $C(f)$ are denoted by crosses.
    Measured values are derived from magnetic injections which are seen in both the \ac{LHO}
    magnetometers and the gravitational wave strain channel, while upper limits on the coupling
    were derived from injections seen by the magnetometers, but not the gravitational wave
    strain channel.
    None of the injections displayed produced measured values for $C(f)$ above 145.3~Hz. 
    The date of each measurement is listed in the legend.
    The time dependence and resonance features are discussed in detail
    in the main text.}
\label{fig:magnetic_gwb}
\end{figure}

The top panel of \fref{fig:magnetic_gwb} shows the coherence $M_{12}(f)$ in
\ac{O3} computed by cross-correlating LEMI-120 magnetometers~\cite{lemi}
between \ac{LHO} and \ac{LLO}.
Schumann resonances are clearly visible as peaks in the coherence spectra.
We also observe higher frequency correlations, which are a known
environmental effect caused by
lightning~\cite{EarthElectromagneticEnvironment, ball:2020}.

Magnetic coupling functions are measured at each detector by
generating oscillating magnetic fields with a known frequency and amplitude at
multiple locations around each site and measuring the resulting signal in the
gravitational wave strain channel (see \sref{ss:pem}).
Results from these measurements are found at~\cite{alog:O3PEMInjections} and
shown in the bottom panel of \fref{fig:magnetic_gwb}.
In \ac{O3}, we monitor time dependence of the magnetic coupling function by
performing the same study weekly throughout the observing
run~\cite{alog:WeeklyMagneticInjections, alog:mag_fac}.
The coupling function data shown in the bottom panel of \fref{fig:magnetic_gwb}
also displays resonances at certain frequencies.
These may be related to resonant motion of optics in the interferometers.
Correlation between the coupling function resonances and \ac{OSEM} channels is
reported in~\cite{alog:MagneticCouplingResonances,alog:MagneticCouplingResonancesFollowUp}.

\section{Data quality for transient gravitational wave analyses}\label{s:transientDQ}

One of the two main classes of analyses for gravitational waves
is focused on identifying and interpreting gravitational waves 
from short duration signals.
As of \ac{O3}, the only source of detected gravitational waves is from \ac{CBC} signals. 
These gravitational wave sources, along with other sources of short
duration gravitational waves, such as supernovae~\cite{Abbott:2019pxc} 
or cosmic strings~\cite{Abbott:2017mem},
are collectively referred to as gravitational wave \emph{transients}. 
This section outlines the main data quality products used for 
transient analyses in \ac{O2} and \ac{O3}, 
data processing required to remove loud glitches, 
data quality methods for low latency identification of gravitational waves, 
and validation of candidate gravitational wave events.  
This includes searches for gravitational waves from  \ac{CBC} 
sources~\cite{Usman:2015kfa,Sachdev:2019vvd,Adams:2015ulm,
QiChuThesis,Venumadhav:2019tad}
and from 
burst sources~\cite{Klimenko:2015ypf,Lynch:2015yin},  
as well as additional analyses completed
to understand the properties of the gravitational wave 
source~\cite{Veitch:2014wba,Ashton:2018jfp}.

\subsection{Low Latency Data Quality}\label{ss:low_lat_dq}

An important component of transient analysis is the detection and distribution
of candidate gravitational wave signals as soon as possible to facilitate multi-messenger
follow up~\cite{LIGOScientific:2019gag}.
In order to support these analyses, data quality products must be available
with the the same or lower latency than the identification of candidates.
In \ac{O2} and \ac{O3}, this was accomplished with data quality products
specifically designed to be used by these low latency searches and human
vetting of candidates.
Additional details on these procedures used in human vetting of low latency candidates
and data quality information that is released alongside public alerts
can be found in \sref{s:validation}.

The first data quality products are produced and
distributed directly alongside
the calibrated gravitational wave strain data.  These include two data streams, 
the \emph{state vector} and the \emph{data quality vector} (data quality vector).  
The state vector is a time-series (recorded at 16~Hz) which contained
information about the overall operating state of the detector, the presence of any hardware
injections, and the status of the online calibration pipeline encoded
as a bit-vector.  The operating status information was limited to an indication that the interferometer was
ready to take observation-quality data, that the site operators had transitioned to an intended
observing mode, and that both the data was of observation quality and the
calibration of the data was successful.  This information could then immediately be used by search
pipelines to know which data should be processed, 
while the information on hardware injections~\cite{Biwer:2017}
could be used to excise these simulated events from analysis results searching for true gravitational waves.
Finally, the additional information describing the state of the calibration could be used to track the
performance of the calibration code, but was designed to be extraneous for the search
pipelines themselves.

The data quality vector is a time-series (also recorded at 16~Hz) which contained
information about the presence of well-known noise sources in the data, again encoded
as a bit-vector.
Because the data quality vector was distributed at the same time as the calibrated
strain,
it was able to be made available with effectively zero latency, but was limited in scope.
This vector indicated the presence of several types of interferometer control-loop
saturations.  A prominent example of these was the glitch located very close temporarily
 with, and subsequently subtracted from, GW170817
in LLO~\cite{GW170817, Abbott:2018wiz, Pankow:2018qpo}.
These saturations occur when control actuation signals rail against software
limits,
thereby producing short-duration, broadband excess noise.
Importantly, the data quality vector, like the data quality flags described 
in \sref{s:flags},
only encoded binary state information, and several searches used the
data quality vector to
automatically reject candidates~\cite{Messick:2016aqy, Nitz:2018rgo}.

In addition to the state and data quality vectors, an additional data product,
the \idq ~\cite{Essick:2020qpo} timeseries, was generated in low latency in 
\ac{O2} and \ac{O3}.
\idq is a statistical inference framework which identifies non-Gaussian noise
in the gravitational wave strain based on
auxiliary channels and produces probabilistic data quality information.
Multiple \idq timeseries are available which have different interpretations,
including $p(glitch|aux)$
and log-likelihood, but all reflect the degree of non-Gaussian noise in the
data. $p(glitch|aux)$
gives the probability that there is a glitch in the gravitational wave strain channel,
given auxiliary channel information, $aux$.
The log-likelihood, in contrast, gives the likelihood ratio between the glitch
and clean models.
$p(glitch|aux)$ is computed from the log-likelihood and also contains prior
odds folding in the glitch
rate within the detector.

In \ac{O3}, candidates were released publicly for multi-messenger follow up
without initial human vetting through \gwcelery~\cite{gwcelery, 
emfollow_public_docs}.
These public alerts were distributed via \ac{GCN} with latencies as low as a few
seconds after merger.
At such a low latency, few data quality products are available:
the aforementioned state and data quality vectors, and the 
\idq timeseries detailing
statistics such as $P$(glitch) and log(likelihood) 
(\idq is discussed further in \sref{s:flags}).
\gwcelery included a detector characterization task which was required to 
be passed before the distribution of any public alert.
This task checked the values of the state vector for
a short window of a few seconds centered about the gravitational-wave
candidate's event time.
The check would pass only if the detectors were in observing mode and no
hardware injections were found.
The information available from \idq $P$(glitch) was not a part of these automated checks, 
but was reported for human inspection.

Early in \ac{O3}, \gwcelery also checked the data quality vector to ensure there
were no overflows,
but this check was removed once all search pipelines switched to using 
gated data (see \sref{s:gating})
to avoid vetoing an overflow that
would have been gated out prior to the search.
All candidates passing this check were assigned the "DQOK" label in 
\gracedb. 

Moving forward, further automation of low latency data quality is planned for
the \ac{O4}
and beyond (see \sref{s:O4expectations}).
Further integration of \idq into the automated stack of data quality checks
and within low latency search pipelines themselves, is a priority,
especially as search pipelines approach and achieve negative latency 
where binary inspirals are detected before merger (i.e., early-warning
alerts~\cite{Sachdev:2020lfd,Kapadia:2020kss,Nitz:2020vym,Magee:2021xdx})
for \ac{O4}.

\subsection{Data Quality Products}\label{s:flags}

As a part of searches for gravitational wave transients, multiple types of data quality 
products are used to indicate the state of the detector and the analyzed data, 
with 
the goal of increasing the sensitivity of the searches and reducing the number 
of 
false alarms. 
In addition to the data quality products available in low latency, 
available data quality products include data quality flags, lists 
of time segments
identifying specific data quality concerns, 
and higher latency predictions of the data quality from \idq. 

When a period of time demonstrating excess noise has been identified along
with an auxiliary channel that witnesses the source of the noise, 
the associated time periods of gravitational wave 
strain data are 
marked as likely to contain instrumental artifacts using a data quality flag. 
A list of data quality flags 
that indicate 
periods of noisy data is then provided to astrophysical searches. 
For \ac{CBC} 
searches, two tiers of 
data quality flags are produced: category~1 and category~2.
An additional tier of data quality flags, category~3, 
was used by some unmodeled transient searches.
When data quality flags used by the searches are referred to as
daat quality vetoes.

Category~1 vetoes indicate that the data have been severely impacted by noise 
and should not be 
analyzed at any stage of an astrophysical analysis.
These problems can indicate either challenges for the searches due to 
significant
changes to the properties of noise in the detectors
or gravitational wave strain data that is not correctly calibrated.
Category~1 vetoes are defined using segments stored in the DQSEGDB 
(see \sref{ss:dqsegdb}), and hence the start and end times
must be integer seconds. 
However, many category~1 vetoes last multiple hours. 
Reasons for category~1 vetoes include incorrect detector configurations,
data dropouts, and on-site maintenance work~\cite{150914_dqvetoes}.
In a subset of these cases, operator logs are used to define the 
category~1 segments instead of auxiliary channel witnesses. 
Data from segments flagged by category~1 vetoes are not ingested by 
astrophysical analyses.
Because category~1 vetoes prevent the detection of gravitational-wave
candidates from the time period they flag, 
this veto category is not used unless a data segment is not 
analyzable. 
In \ac{O2} and \ac{O3}, category~1 vetoes removed less than $2.0\%$ of 
analyzable 
data at each detector. 

\begin{figure}[ht]
    \centering
    \includegraphics[width=\textwidth]{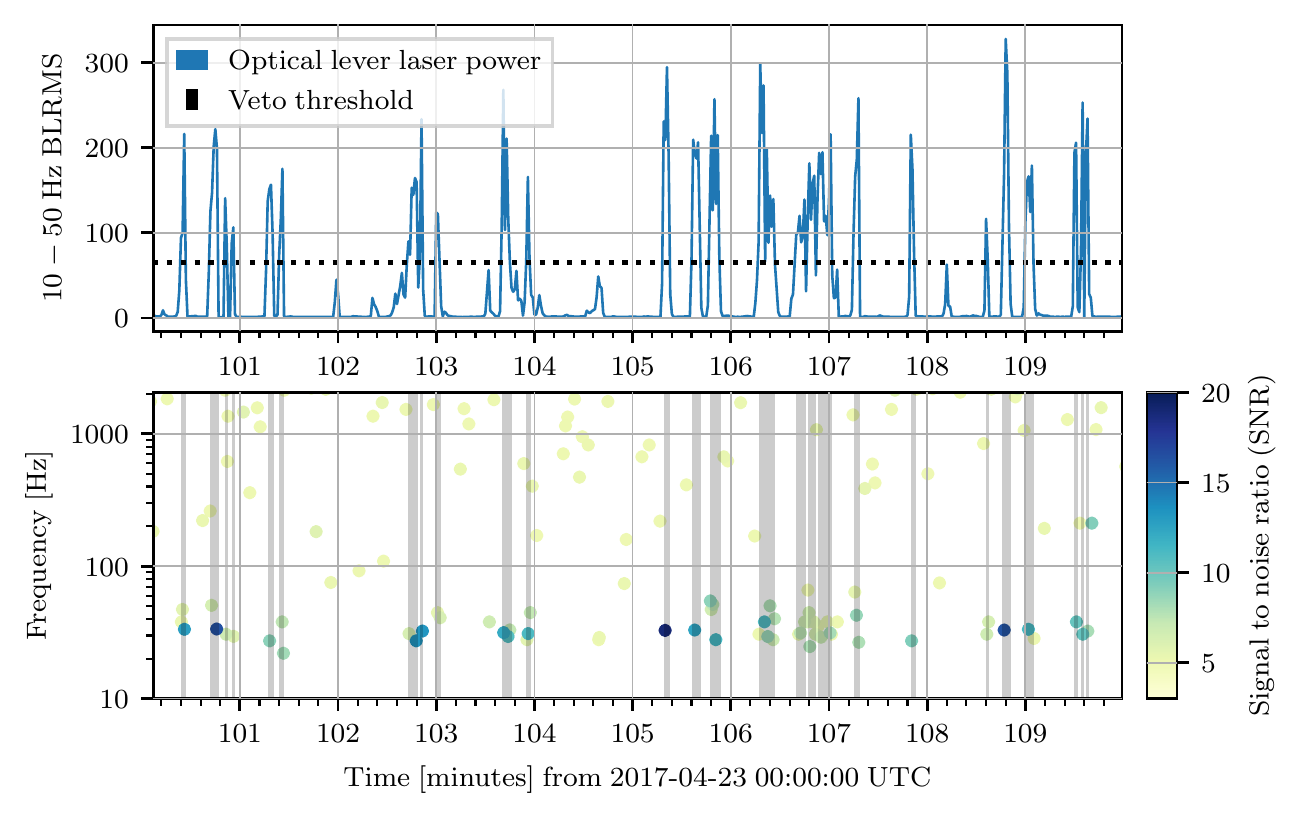}
    \caption{A data quality veto developed in \ac{O2} based on the power in an 
optical lever laser. 
    \textit{Top:} 10-50~Hz band-limited RMS of the optical lever laser power as 
a function of time. 
    The dotted black line indicates a threshold value above which a data 
quality veto is generated. 
    The threshold value is determined based on how efficiently a given 
threshold removes transient noise 
    from the data.
    \textit{Bottom:} A time-frequency scatter plot of transient noise triggers 
on the same time scale. 
    The color bar indicates the signal-to-noise ratio of each trigger. The 
shaded regions indicate the 
    presence of a data quality veto. Any trigger within a shaded region is 
considered to be instrumental 
    noise and is recommended to not be considered a potential candidate
by astrophysical search algorithms. 
A population of glitches 
    with \fixme{\ac{SNR} $>$ 8} and peak frequency between 20-80~Hz is targeted 
and removed by this veto.
    }
    \label{figure:veto-example}
\end{figure}

Category~2 vetoes indicate periods of time where the data is impacted by excess 
noise
and should be treated with caution, but can still be used as input to an 
astrophysical analysis. 
Potential candidates that are identified during
data flagged as category~2 are recommended to not be considered potential candidates by 
astrophysical search pipelines
as candidates identified during this time have a higher likelihood to have been 
caused by 
instrumental artifacts than candidates identified outside category~2 periods.
Category~2 vetoes are also defined using integer second segments in the
same manner as categeory~1 vetoes. 
However, each category~2 segment generally only a few seconds in duration.
\Fref{figure:veto-example} shows an example of a category~2 veto that was 
developed during \ac{O2}. 
This veto was used to flag periods of excess noise in an optical lever channel 
at \lho 
that produced glitches in the gravitational wave strain data. 
Removing these short time periods removed a significant portion of the glitches
with \ac{SNR} $> 8$ during this data stretch. 
Other examples of category~2 vetoes include earthquakes, thunder, 
and high wind conditions~\cite{150914_dqvetoes}.
Information available as a part of the data quality vector is also used 
to create data quality flags that ingested by high latency searches 
as category~2 vetoes. 
Since category~2 vetoes also reduce the duration that a search pipeline is able to 
detect a 
gravitational wave event, utilization of these vetoes risks reducing the total number of 
detectable signals if the 
veto is not properly tuned or the amount of time removed is too large.
For these reasons, 
new category~2 vetoes are not used by searches unless the flagged data quality issue
has been shown to severely impact the search, 
minimizing  
the amount of time removed by category~2 vetoes as much as possible.
In \ac{O2} and \ac{O3}, \ac{CBC} category~2 vetoes removed less than $0.4\%$ 
of analyzable 
data at each detector.
The total amount of time removed by each category for all observing periods
is shown in \tref{tab:veto_sum}.

\begin{table*}[t]
        \begin{tabularx}{\textwidth}{l@{\extracolsep{\fill}}rrrr}
        \textbf{Detector} & \textbf{CAT1} & \textbf{CBC CAT2} & \textbf{Burst 
CAT2} & \textbf{Burst CAT3} \\
        \hline
        
\makebox[0pt][l]{\fboxsep0pt\colorbox{lightgray}{\mystrut\hspace*{0.97\linewidth
}}}LIGO Hanford O2 & \LHOCATONEOTWO \% & \LHOCATTWOCBCOTWO \% & 
\LHOCATTWOBURSTOTWO \% & \LHOCATTHREEBURSTOTWO \% \\
        LIGO Livingston O2 & \LLOCATONEOTWO \% & \LLOCATTWOCBCOTWO \% & 
\LLOCATTWOBURSTOTWO \% & \LLOCATTHREEBURSTOTWO \% \\
        
\makebox[0pt][l]{\fboxsep0pt\colorbox{lightgray}{\mystrut\hspace*{0.97\linewidth
}}}LIGO Hanford O3A & \LHOCATONEOTHREEA \% & \LHOCATTWOCBCOTHREEA \% & 
\LHOCATTWOBURSTOTHREEA \% & \LHOCATTHREEBURSTOTHREEA \% \\
        LIGO Livingston O3A & \LLOCATONEOTHREEA \% & \LLOCATTWOCBCOTHREEA \% & 
\LLOCATTWOBURSTOTHREEA \% & \LLOCATTHREEBURSTOTHREEA \% \\
        
\makebox[0pt][l]{\fboxsep0pt\colorbox{lightgray}{\mystrut\hspace*{0.97\linewidth
}}}LIGO Hanford O3B & \LHOCATONEOTHREEB \% & \LHOCATTWOCBCOTHREEB \% & 
\LHOCATTWOBURSTOTHREEB \% & \LHOCATTHREEBURSTOTHREEB \% \\
        LIGO Livingston O3B & \LLOCATONEOTHREEB \% & \LLOCATTWOCBCOTHREEB \% & 
\LLOCATTWOBURSTOTHREEB \% & \LLOCATTHREEBURSTOTHREEB \% \\
        \hline
        \end{tabularx}
        \caption{Percent of single-detector time removed by each category of 
veto for each detector.
                }
        \label{tab:veto_sum}
\end{table*}

Unlike \ac{CBC} searches, unmodeled burst transient searches cannot rely on a 
specific chirp-like morphology for the gravitational wave signal, which 
increases
the difficulty of rejecting background noise compared to a matched-filter 
search. 
For these unmodeled searches, additions and extensions are made to the
category~2 definitions used in \ac{CBC} searches, resulting in an
 increase in time removed relative to the \ac{CBC} definitions.  
While multiple unique category~2 flags
were added for standard burst vetoes in \ac{O2} and \ac{O3}, the majority of this increase 
in vetoed time was
 due to flags removing very loud glitches (typically \ac{SNR} of $>$100) at 
both interferometers
associated with light intensity dips. 

For some unmodeled transient searches, e.g., 
\cite{PhysRevD.100.024017,Abbott:2016tdt},
category~3 data quality flags are also applied as a final stage after initial 
triggers are
generated.  These flags are mostly produced by the \hveto~\cite{Smith:2011an} 
algorithm 
(see \sref{s:noiseinvestigations}), which 
is run over strides of about five days total coincident observing of both LIGO 
interferometers to automatically identify auxiliary channels which can be
used to create statistically significant vetoes.  
Using predetermined thresholds
on veto efficiency and statistical significance, as well as durations and 
\ac{SNR}
thresholds selected by the \hveto algorithm for each channel, an
additional set of data quality flags are constructed.  Unlike category~1 and 2,
category~3 vetoes may be of sub-second duration rather than being defined using
integer seconds.  

Targeted analyses, such as those investigating potential gravitational wave counterparts 
to 
gamma ray bursts~\cite{Abbott:2020yvp} or fast radio 
bursts~\cite{others:2016ifn}, 
also utilize data quality products. 
These targeted analyses use the same data quality products as non-targeted 
searches with similar 
search techniques. 
Targeted searches using matched filtering rely upon \ac{CBC} data quality 
products, 
while unmodeled targeted searches use burst data quality products. 

The segments for each category of veto used in analyses of LIGO data are available 
via \ac{GWOSC}~\cite{Abbott:2019ebz,gwosc_o1_strain,gwosc_o2_strain}.
Veto segemnts for \ac{CBC} and Burst searches are provided separately. 
Alongside the categories discussed in this work, 
segments corresponding to time periods where hardware injections are 
underway are also released. 

To measure the effects of data quality vetoes on astrophysical searches,
the PyCBC search pipeline~\cite{Usman:2015kfa}
was run on \ac{O3} data with and without using category~2 data quality vetoes
to remove data.
Data from both \ac{LHO} and \ac{LLO} recorded between April 18, 2019 and 
April 26, 2019 was used in this comparison. 
In both cases, the PyCBC pipeline was ran with the exact same configuration as used in 
\cite{GWTC-2}, but without including category~2 vetoes for one analysis. 
During this analysis period, the only category~2 vetoes that existed
were at \ac{LLO}, so all additional discussion will focus on the impact of 
these vetoes on the \ac{LLO} data or the combined \ac{LLO} and \ac{LHO}
analysis. 
These category~2 vetoes specifically vetoed noise that was due to acoustic
from thunder and vibrations from a camera shutter that was inadvertently
left operating on a timer while the detector was in observing mode. 

\begin{figure}[ht]
    \centering
    \includegraphics[width=\textwidth]{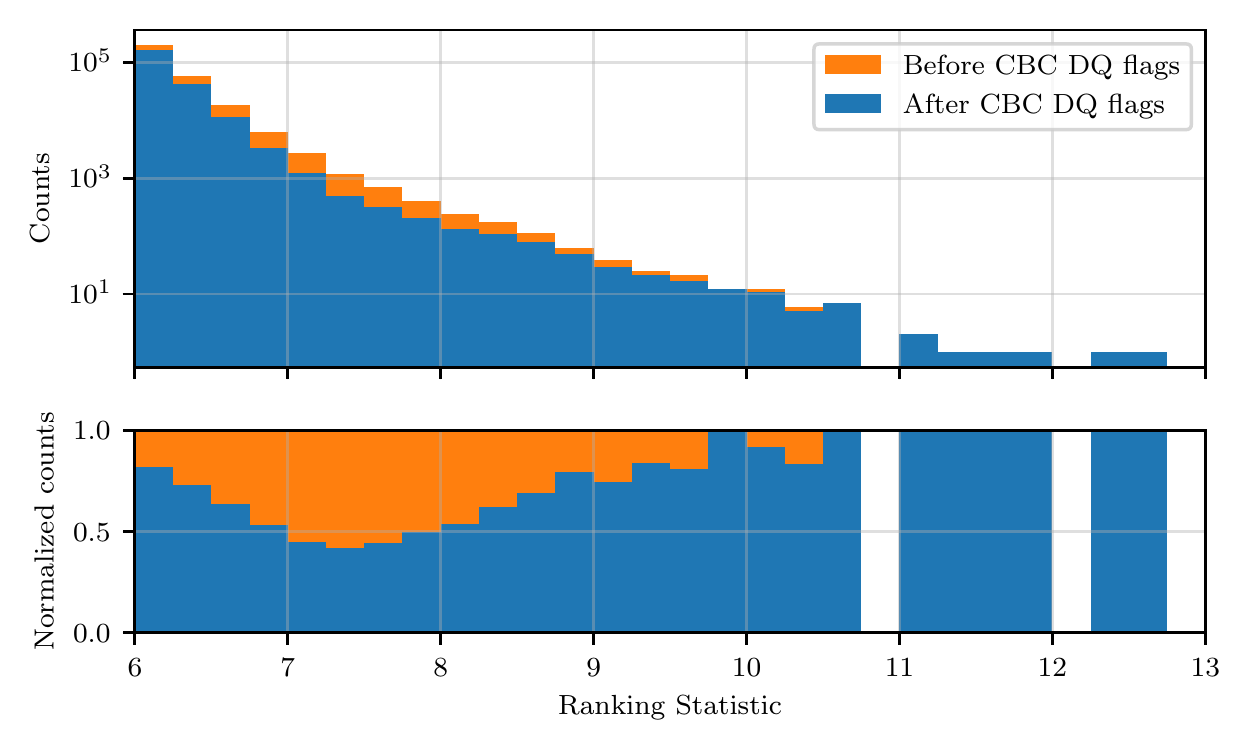}
    \caption{A comparison of the total number of triggers considered by PyCBC
    in the LIGO Livingston data before and after including 
    \ac{CBC} category~2 data quality flags.
    Triggers are binned by the assigned PyCBC ranking statistic.
    \textit{Top:} A histogram of the total number of triggers identified
    by PyCBC in each ranking statistic bin before and after including data
    quality flags.
    \textit{Bottom:} A histogram of the same triggers shown on top, 
    but now normalized so that that total number of triggers before including
    data quality flags is $1$. 
    The difference between the before and after height is the
    fraction of triggers in that bin removed by data quality flags. 
    }
    \label{figure:pycbc-hist}
\end{figure}

Category~2 vetoes are designed to remove a higher fraction of triggers
than the fraction of time that is removed. 
Using these two PyCBC analyses of the same time period, we can directly
test if this is true. 
All triggers recorded by PyCBC are ranked based on the recovered
matched filter \ac{SNR} and multiple signal consistency tests.
This ranking is referred to as the ``ranking statistic.''
After binning triggers based on the recovered ranking statistic, 
we compare the total number of triggers before and after the inclusion
of category~2 data quality vetoes. 
This comparison, both in terms of the total number of triggers and
the fractional differences in the number of triggers, 
is shown in \fref{figure:pycbc-hist}. 
Only $0.78\%$ of data was removed by category~2 vetoes
during this analysis period, 
but over $50\%$ of triggers were removed for some ranking statistic bins. 
Considering all triggers with a ranking statistic above 6.5, 
$42\%$ of triggers were removed.
The large difference in the fraction of time removed by the data quality flags
and the fraction of triggers removed shows that the data quality flags
are removing time periods that produce a large number of triggers in the 
PyCBC search, as designed. 
The specific ranking statistic bins that have the highest fraction 
of triggers removed by the category~2 vetoes during this analysis are 
consistent with the amplitude of the noise that was vetoed. 
The data quality issues present during this analysis were particularly
problematic for the analysis because they were loud enough to create
significant background triggers but quiet enough to not
be easily rejected by signal consistency checks.  

\begin{figure}[ht]
    \centering
    \includegraphics[width=\textwidth]{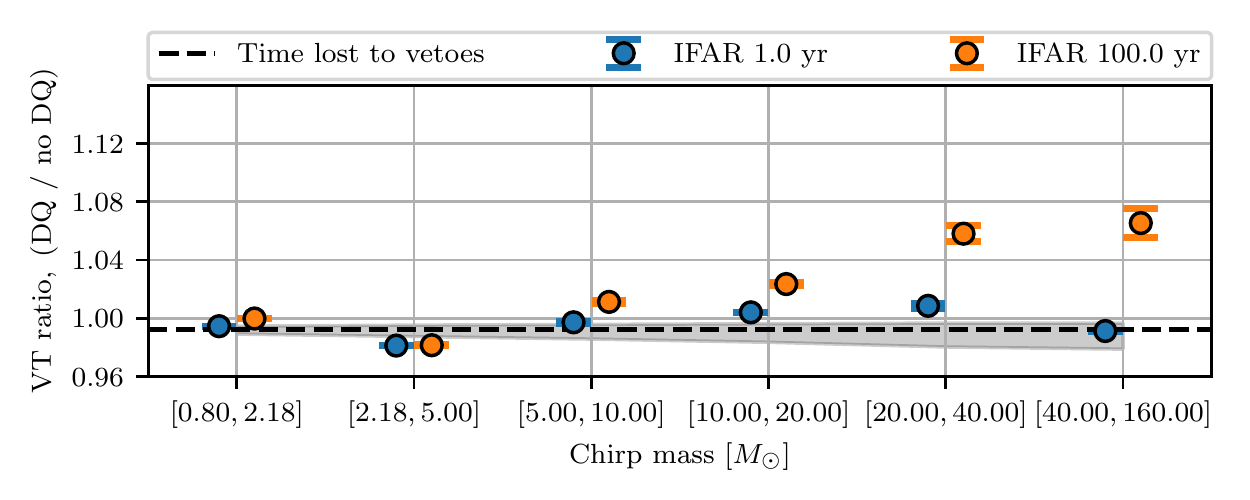}
    \caption{Ratio of sensitive volume-time for several chirp mass bins as 
    measured by the PyCBC search 
    pipeline before and after the inclusion of category~2 data quality flags. 
    The dotted line indicates the fraction of time removed by data quality flags,
    with the shaded grey region showing the $1\sigma$ Poisson error in the 
    total number of injections that are expected to occur during this time.
    The blue points show the ratio of the measured volume-time (VT) in each chirp mass
    bin at a significance threshold of 1 per year, 
    while the orange points used a significance threshold of 1 per 100 years. 
    Error bars on each point represent the $1\sigma$ error in the measured 
    volume-time ratio after taking into account correlations between the set of simulated
    signals found before and after data quality flags are included.
    For lower mass bins, the average measured sensitivity decreases 
    slightly, consistent with 
    removing approximately 1\% of analyzable time from the search pipeline. 
    In the two of the highest mass bins, the 
    average sensitivity increases by approximately $5\%$ when using a 
    significance threshold of 1 per 100 years.
    }
    \label{figure:vt-ratio}
\end{figure}

In addition to considering the total number of triggers identified by PyCBC, 
we also compare the sensitivity of the PyCBC search to astrophysical signals
before and after the inclusion of category~2 data quality flags. 
The sensitivity of PyCBC is assessed by analyzing a large number
of simulated signals with the analysis pipeline, and
measuring the efficiency with respect to distance
at which these signals are identified by the pipeline. 
The distance at which the pipeline recovers $50\%$
of simulated signals is referred to as the ``sensitive distance.'' 
The volume of a sphere with this distance as a radius,
multipled by the total amount of time analyzed is called the 
volume-time.
The measured volume-time of an analysis is the metric that most directly
estimates the total number of real gravitational-wave signals
that a pipeline could detect. 
We measure the volume-time of PyCBC before and after the 
inclusion of category~2 data quality flags. 
The ratio of these measured volume-time at two different significance
levels, is shown in \Fref{figure:vt-ratio}. 
Since different regions 
of the parameter space are more susceptible to instrumental artifacts, the 
sensitivity measurement is broken 
up into several chirp mass bins. 

Overall, there does appear to be a modest increase in the measured volume-time
of PyCBC after including data quality flags.
For the lower chirp mass bins, sensitivity 
does not change at a statistically 
significant level if noisy data are removed. This is expected, given that low 
mass systems produce longer 
duration waveforms and signal consistency tests are robust in this 
region of the parameter space~\cite{TheLIGOScientific:2017lwt}. 
The higher mass regions typically produce shorter duration waveforms, 
$\mathcal{O}$(0.1 - 1 seconds), and can have a 
significant fraction of their template waveform overlap with a noise transient. 
Removal of noisy data 
results in a 5\% increase in volume-time in the highest chirp mass bins.
This increase is lower than observed in the
\ac{O1}~\cite{TheLIGOScientific:2017lwt},
likely due to improvements to the PyCBC 
pipeline~\cite{Nitz:2017svb,Nitz:2017lco}.
The impact on the volume-time of other \ac{CBC} searches is expected
to be comparable to the impact on PyCBC. 

In addition to data quality flags, the \idq ~\cite{Essick:2020qpo} timeseries
was used as a part of the GstLAL search pipeline 
in \ac{O3}~\cite{Sachdev:2019vvd,Godwin:2020weu}. 
The generation of the higher latency \idq dataset was identical to that of 
low latency \idq, but the \idq model was retrained using the 
full dataset available.
Investigations into the impact of inclusion of \idq 
information into the 
GstLAL search pipeline
have also shown increases in search sensitivity~\cite{Godwin:2020weu}.

\subsection{Gating for transient searches}\label{s:gating}

In addition to data quality products that are used in post-processing and significance 
calculations 
of transient searches, other data quality issues must be addressed by a 
pre-processing step.
In \ac{O2} and \ac{O3}, a process called ``gating'', in which the data were 
multiplied
by a smooth inverse window function, was used to remove high amplitude
noise transients from gravitational wave strain data.
If a noise transient has sufficient amplitude, it will ring up the whitening 
filters and imprint 
the impulse response of 
those filters onto the data, corrupting several seconds of 
data on either side of the transient~\cite{Usman:2015kfa}. 
Therefore simply discarding candidates that are from times impacted
by these artifacts, as is done using category~2 and 3 veotes, 
is not sufficient to mitigate their impact. 
Since these transients are often sub-second in duration and the region of 
corrupted data is often 
$\mathcal{O}$(10) seconds in duration, it is advantageous to be able to remove the 
sub-second glitch before the 
whitening filter is applied, recovering the surrounding data. 
The gating process multiplies the data by an 
inverse Tukey window which rolls smoothly from 1 to 0, zeroing out the data 
containing the large transient 
and leaving all data outside of the window unchanged~\cite{GW170817}. 
This procedure was used to mitigate the impact of the glitch near 
GW170817 before searching the data~\cite{GW170817,Pankow:2018qpo}

\begin{figure}[ht]
    \centering
    \includegraphics[width=\textwidth]{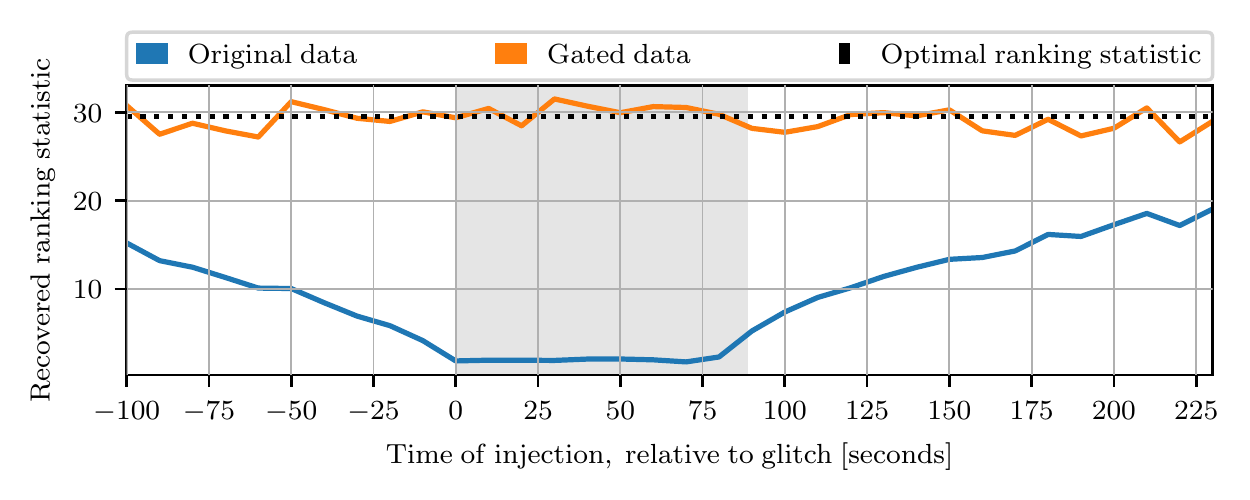}
    \caption{A comparison of recovered ranking statistic with gated 
    and un-gated gravitational-wave strain data.
    The shaded region indicates the time period when the simulated signal 
    would directly overlap with the loud glitch in the data. 
    The ``optimal ranking statistic'' is the expected ranking statistic
    that this simulated signal would be assigned in stationary, Gaussian data. 
    In the case where no gating is used, the assigned ranking statistic is 
    significantly reduced as compared to the optimal ranking statistic.
    }
    \label{figure:gating}
\end{figure}

As an example of how this procedure allows for astrophysical signals
to be detected by search pipelines despite the presence of a nearby loud
glitch, we performed a series of injections near such a loud glitch,
before and after gating, and measured the recovered ranking statistic
for each injection.
Simulated signals from the merger of two non-spinning $1.4 M_\odot$ 
components were injected near the loud glitch at GPS time 1253878751
in gravitational wave strain data from \ac{LLO}.
Data from the surrounding 1024 seconds was used to calculate the 
power spectral estimate of the data. 
Each injection was then recovered using the PyCBC single-detector
ranking statistic~\cite{Usman:2015kfa}, 
before and after mitigating the glitch with the previously
described gating method. 
The result of this study is shown in
\fref{figure:gating}.
The recovered ranking is consistent with that expected with Gaussian noise
after gating the loud glitch.
Without gating, the simulated signal is recovered at a much lower ranking statistic
than expected, even for time periods where the signal does not overlap the glitch. 
When the simulated signal overlaps the glitch, the recovered ranking statistic is
reduced due to signal-consistency tests that are used by PyCBC to reject 
instrumental artifacts. 
The presence of the loud glitch also biases the estimate of the 
power spectral density, which reduces the recovered ranking statistic
during the entire period considered.  

In \ac{O2} and \ac{O3}, a subset of instrumental issues that 
were correlated with high amplitude transient noise were indicated to \ac{CBC} 
searches using gating windows rather 
than data quality vetoes. While most of these glitches were sub-second in 
duration, they sometimes 
manifested at a high enough rate that consecutive gating windows were 
constructed, leading to several 
seconds of data being set to zero. To avoid underestimating the steady state 
noise in the detector, 
any gating window longer than 3 seconds in duration was considered a category~2 
veto. In addition, 
several search pipelines implemented a complementary automatic gating procedure 
that constructed 
gating windows to remove transient noise above a set threshold on the whitened 
data amplitude
 ~\cite{Usman:2015kfa,Sachdev:2019vvd}.

Additional methods to remove glitches exist that can more precisely remove 
glitches
at the cost of additional computational 
complexity~\cite{Cornish:2014kda,Cornish:2020dwh,Zackay:2019kkv}.
While gating has been shown to bias post-detection estimation of gravitational 
wave source 
properties~\cite{Pankow:2018qpo}, 
when used sparingly, 
gating with the windowing method discussed in this section does not measurably reduce 
the sensitivity of \ac{CBC} searches, beyond what is expected from the 
amount of data removed by this process.

Similar gating methods are utilized for continuous searches, which calculate 
the noise
spectrum over longer periods than in transient analyses, and hence are more 
susceptible
to high-\ac{SNR} glitches.
The impact and methods used for these searches are discussed in section \ref{s:longdurDQ}.

\subsection{Unvetoed noise sources}

Despite the wealth of information available from the LIGO summary pages and 
various
automated algorithms to  correlate noise with auxiliary channels, several 
classes
of glitches have persisted in the data with insufficient clues to remove them 
all. 
These glitches can be 
categorized based on their time-frequency evolution. Three main types of 
glitches that 
have been harmful to transient searches are short duration transients, 
scattered light, and extremely loud
glitches~\cite{LIGOScientific:2018mvr,TheLIGOScientific:2017lwt}.
 
The two main classes of short duration transients that impact the searches
are blips and tomte glitches. 
Blip glitches (see \sref{ss:blips}) 
are particularly bad for searches for high mass binary black holes and 
generic 
transients~\cite{GW150914_detchar,Davis:2020nyf,TheLIGOScientific:2016uux}. 
The short duration makes it difficult to differentiate 
a short astrophysical signal from a blip.
Although they have not appeared in multiple detectors in coincidence
more than expected due to chance, 
they create outliers in time-shifted backgrounds. 
Similar issues have been noted for tomte glitches.
While blip glitches generally have power present at frequencies above the merger
frequency of high mass \ac{CBC} signals, tomte glitches are generally lower in 
frequency
and better line up with a high-mass merger template. 
This morphology means that signal consistency tests designed to address the 
impact of blip
glitches~\cite{Nitz:2017lco} are not as effective for tomte glitches. 

Unmodeled burst searches make use of parameter space bins in order to contain problematic 
glitches that cannot be otherwise vetoed.   
One such analysis pipeline that we will consider is the 
coherent wave-burst (cWB) search~\cite{Klimenko:2015ypf}.
Sorting candidates that are similar to known glitches into seperate bins and 
then measuring the background of each bin seperately 
increases the sensitivity of the other bins while still exploring the entire parameter 
space.  The cWB search is especially sensitive 
to the blip class of glitches.
These glitches are identified by cWB using the measured quality factor ($Q$),
which is the measured ratio of the energy contained in the identified candidate 
to the background energy from nearby data. 
Bin 1, which contains the majority of blip glitches,  
includes triggers that
have a quality factor $< 3$ and bin 2 contains the remaining triggers 
~\cite{Abbott:2016ezn,PhysRevD.100.024017}.  Before \ac{O3} this was sufficient to contain 
the blip glitches.  During \ac{O3}, exceptionally loud and short ($~$ms) triggers were found 
polluting bin 1 enough to create a new bin to contain them.  A new bin of triggers with 
negligible energy outside their short pulse core are separated into bin 1a while the remaining 
bin 1 triggers are placed in bin 1b ~\cite{shortBurst}.  Bin 2 contains the rest of the 
triggers and had the cleanest background distribution.

Another type of noise that frequently impacts searches for gravitational wave transients 
is scattered light. Scattered light (see \sref{ss:scatter})
especially hurts \ac{CBC} searches for high spin signals, where the glitch
morphology is similar to these types of systems, 
and long duration signals, where the chance of overlap with a scattering 
glitch is high.
Even when the source of scattering is known, the large amount of time that is impacted
by this type of glitch makes it difficult to design an effective data quality 
flag
without removing significant amounts of data and reducing the volume-time of the
searches.

Extremely loud glitches in \ac{O3} were also a severe problem for searches. 
While these glitches are generally removed in transient searches with gates, 
there are cases where glitching correlated with the extremely loud glitch, 
but not as loud as the main glitch, is present. 
These additional glitches have been shown to be a significant contribution to 
the 
background of transient searches. 

There also exist data quality issues that are only partially addressed due to the limitations
of current data quality products.
Since data quality vetoes are designed to maximize the ratio of efficiency 
to deadtime, 
a known data quality problem may not be entirely removed if the efficiency 
versus
deadtime is not high. 
This was the case for weak optical lever glitches in O2.
An additional limitation of the process currently used to egenrate data quality products 
is that data quality flags are tuned
to remove excess power from glitching, as opposed
to only targeting time periods that are problematic for the search pipelines. 
This procedure reduces the risk of unintential bias in generating data quality flags, 
but also limits their effectiveness. 

The combined impact of these unvetoed noise sources varies by analysis 
pipeline.  As an example, we 
conducted a short study on the impact of unvetoed glitches on a \ac{CBC} search 
for gravitational wave events 
coincident with externally detected \ac{GRB} events.  This study found that 
of high \ac{SNR} injections in the \ac{LLO} gravitational wave strain data
that were not found by the search pipeline,
approximately 35\% were missed due to 
contamination by one of the 
categories of noise sources listed above.  These missed injections are tests of 
the pipeline's ability to
detect a simulated (injected) signal within the data, and missed high \ac{SNR} 
injections indicate that the
analysis's ability to detect gravitational waves is negatively affected by these noise 
sources.  The noise directly 
reduces the effective astrophysical distance of gravitational waves to which the search is 
sensitive,
typically defined for this type of targetted search as the distance at which $90\%$ of injections
are successfully recovered~\cite{Abbott:2020yvp}.  
The missed injections considered in this study are at distances consistent 
with the $90\%$ recovery distance, meaning that these noise sources are one of the limiting
factors to the sensitivity of the \ac{GRB} search. 
Thunderstorm activity, as described in \sref{thunderstorms}, had an additional 
impact in causing 
extended time segments with elevated background triggers in the searches.  
These impacts may be mitigated through application of additional 
search-specific 
vetoes, through additional padding of existing vetoes, and through additional 
identification of strong 
interferometer witnesses of the noise sources in the future.

\subsection{Validation Procedures}\label{s:validation}

An essential element of gravitational wave detection is the process of evaluating the 
impact of
data quality issues on candidate events,  
known as event validation~\cite{GWTC-2}.
Event validation serves to both increase confidence in the astrophysical origin 
of gravitational-wave signals by investigating the possibility the event is
instrumental in origin, as well as increasing confidence in analyses of events
by evaluating the impact of any relevant data quality issues. 
These investigations were an important component in confirming the first 
detected gravitational wave event~\cite{GW150914_detchar}, 
and continue to be fruitful as the rate of detection 
increases.

In \ac{O2} and \ac{O3}, event validation was completed on two timescales:
within 30 minutes to multiple days of event 
identification~\cite{LIGOScientific:2019gag}, 
and weeks to months after event identification~\cite{GWTC-2}. 
The first, low latency validation step focused on evaluating detection 
confidence and 
the impact of data quality issues on source classification and sky localization 
in order
to support electromagnetic follow up efforts. 
The high-latency validation step utilized additional tools not available in 
low latency, such as additional statistical data quality metrics, 
and supported analyses of gravitational-wave source properties. 

An important facet of the gravitational wave analyses in \ac{O2} and \ac{O3} 
was the responsibility of data
quality experts to vet candidate gravitational wave signals in low latency as a
part of the rapid response team, which convened to make decisions
about releasing or retracting candidates for multi-messenger 
follow up~\cite{LIGOScientific:2019gag}.
The rapid response team was alerted whenever a significant candidate appeared,
usually within a few seconds, and examined the data quality at the time
of the event.
Low latency searches~\cite{Adams:2015ulm,QiChuThesis,Lynch:2015yin, 
Klimenko:2015ypf, 
Messick:2016aqy, Nitz:2018rgo}
require special attention due to the possibly high scientific cost of missing 
transient events
in real-time or falsely initiating electromagnetic follow up campaigns.

The tools available to make data quality decisions matured significantly
since \ac{O1}. 
In \ac{O2}, only a few data quality products were available in low latency, 
most notably omega scans of auxiliary 
data~\cite{Chatterji:2004qg,Chatterji_thesis,gwdetchar}.
In \ac{O3}, a large number of additional tools were generated in low latency, 
including additional
time-frequency visualizations and monitors
of the gravitational wave strain data~\cite{Zevin:2016qwy,Mozzon:2020gwa,ldvw}
and data from hundreds of auxiliary channels monitoring the detectors and their
environments~\cite{Effler:2014zpa,Essick:2020qpo},
as well as
identification of likely sources of glitches by correlation with
auxiliary channels~\cite{Smith:2011an,Robinet:2020lbf}.
These data quality products were collated as a 
part of the \dqr (DQR)~\cite{dqr}, introduced in \sref{ss:dqr}.

Even including the time required for the rapid response team to review data 
quality
products and make release decisions, significant candidates were usually 
announced
less than 30 minutes after they were identified by the end of 
\ac{O2}~\cite{LIGOScientific:2019gag}.
Updates to automatically generated alerts in \ac{O3} were delivered on a 
similar timescale.
The median reponse time of the rapid response team in \ac{O3} was \RESPONSETIME. 
Based on the decision of the rapid response team, each candidate
in \gracedb was assigned the "ADVOK" label, indicating that 
there was not evidence to support retraction of the candidate, 
or the "ADVNO" label, indicating that there was evidence to support
retraction.  

\begin{figure}[ht]
    \centering
    \includegraphics[width=\textwidth]{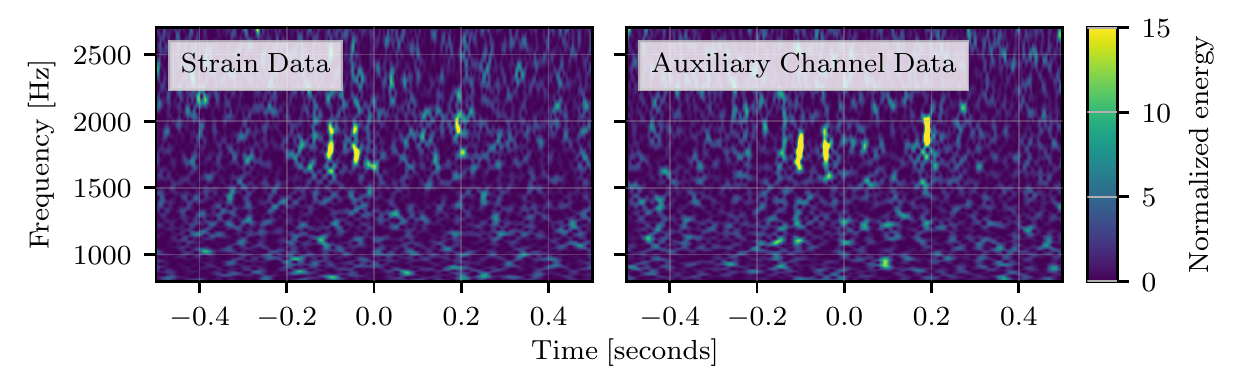}
    \caption{Spectrograms of the gravitational wave strain data from \ac{LHO} during  
    S191110af~\cite{GCN26250} and an 
    auxiliary channel that witnessed the source of the excess power. 
    The similar morphology of the strain and auxiliary channel data suggested that
    the excess power from S191110af was not astrophysical in origin
    and was instead due to an instrumental artifact whose source was in the
    output mode cleaner.
    The association between the gravitational wave strain data and this auxiliary channel was confirmed
    to be statistically significant using \hveto.}
    \label{figure:validation}
\end{figure}

These low latency efforts identified numerous cases where retraction of low latency 
candidates was 
required~\cite{GCN24591,GCN25301,GCN26250}, 
as well as additional cases where instrumental artifacts may potentially bias 
estimates of the
sky localization~\cite{GCN24950,GCN25876,GCN26402}.
One example of a retracted candidate,
identified in low latency as a potential burst source,
can be seen in \fref{figure:validation}. 
Visualizations of auxiliary data at the time of the event allowed identification 
of an auxiliary channel that is a witness
to the excess power observed in the strain channel. 
This auxiliary channel, 
a monitor of a piezoelectric driver that is used to dither
one of the mirrors in the output mode cleaner,
suggested that the candidate was not astrophysical in origin, 
and was instead an instrumental artifact.
Additional statistical analyses~\cite{Smith:2011an} of this time period were 
able to confirm that this
correlation was statistically significant,
which led to the retraction of the candidate~\cite{GCN26250}.

In cases where non-observing mode data is potentially used in analyses, 
evaluating this data is an important component of event 
validation~\cite{GWTC-2}.
While this data is generally not usable for astrophysical analyses, 
due to active work on the detector or lack of calibrated data, 
in rare instances this data has been found to be consistent
with the sensitivity and data quality of typical observing 
periods~\cite{GW170608,GW190814}
after significant work to understand the data quality and calibration
at the time of the events.

Event validation is especially important for short duration signals,
where signal consistency tests are less powerful,
and signals with extreme properties, which are more likely to 
be morphologically consistent with glitches. 
For example,
the most significant intermediate mass black hole candidate in \ac{O2} was found to be consistent
with an optical lever glitch, highlighted in 
\fref{figure:veto-example}~\cite{Salemi:2019ovz}.
Conversely, event validation was an important component
of confirming the astrophysical origin of the largest mass \ac{CBC} source, 
GW190521~\cite{GW190521Adiscovery}.
Candidates with high mass ratios and high spins were
a large portion of the marginal candidates in GWTC-1
which were likely due to instrumental artifacts~\cite{LIGOScientific:2018mvr}.

Even when a candidate is astrophysical, validation is important 
to evaluate potential mitigation for
spurious glitches coincident with the candidate.
For cases when glitches were identified, 
glitches were subtracted~\cite{Cornish:2014kda,Cornish:2020dwh,Davis:2019}
or a reduced time duration and bandwidth was used in analyses.
The increased glitch rate in \ac{O3a} as compared to previous 
observing runs, 
along with an increased event rate, led to mitigation steps for \GWTCTWOMITIGATION
events in GWTC-2~\cite{GWTC-2}. 
A similar number of mitigations is likely required for events identified 
in \ac{O3b}.
These event validation efforts are expected to continue to play a significant
role as gravitational wave detectors become more sensitive, 
and the source properties of gravitational wave are probed to higher precision.

\section{Data quality for persistent gravitational wave searches}\label{s:longdurDQ}

Searches for persistent gravitational wave sources include those for
quasi-monochromatic signals from rapidly rotating neutron stars as well
as stochastic backgrounds due to astrophysical or cosmological sources.
Many searches have been carried out using initial LIGO/Virgo data and
using Advanced LIGO/Virgo data, though no persistent signals have yet
been confidently detected~\cite{O2targetedCW,O2narrowbandCW,O2allskyCW,O2scox1CW,O3atargetedCW,O2StochIso,O2directionalStoch}.

Persistent gravitational wave searches are impacted by different types of
detector noise, usually by noise sources that are persistent as well.
Spectral artifacts in detector data, narrow in frequency and with long-term
coherence (called \emph{``lines''}), pose significant challenges in analyses
for persistent, narrowband gravitational wave signals~\cite{Covas:2018}.
Lines are typically caused by external disturbances (e.g., 60~Hz power mains,
suspension resonances, electronic/magnetic coupling, etc.) that appear as
artifacts in the main gravitational wave strain time series.
Broadband artifacts in detector data typically do not degrade searches for
persistent narrowband gravitational wave searches, but coupling of magnetic
fields to the gravitational wave channel may degrade broadband gravitational
wave searches~\cite{SchumannRes} (see \sref{sec:schumann}).
Additionally, frequent, large-amplitude transient glitches
(\sref{sec:loudtriggers}) can impact analyses for persistent gravitational
waves by degrading power spectral density estimation used in these searches
(see \sref{sec:selfgating}).

Narrowband spectral artifacts in detector data can degrade searches in
different ways, either obscuring or mimicking putative astrophysical signals;
analogous to transient noise obscuring or mimicking transient astrophysical
signals.
Detection pipelines may yield many spurious outliers caused by these artifacts
that require laborious follow-up or require raising significance thresholds;
this leads to a lower possibility for detecting a putative persistent
gravitational wave signal.
Alternatively, such artifacts may obscure a real gravitational wave signal if
the two are close enough so that the signal power is obscured.
These two impacts lead to degraded efficiency and sensitivity of analysis
pipelines.

It is therefore advantageous to mitigate---wherever possible---these spectral
artifacts.
Simply ignoring specific frequencies where lines may obscure a putative signal
degrades the overall parameter space searched and increases the likelihood of
overlooking a true signal.
Where it is not possible to mitigate these artifacts, we are forced to identify
the cause of the lines so that an astrophysical signal is not confused with a
spectral artifact.
This is a difficult, iterative process because low-level spectral artifacts
require long stretches of data in order to be visible above the noise (this is
essentially analogous to the long stretches of data that persistent
gravitational wave signal detection analyses use).
Sometimes lines may be spaced equally distant from each other in frequency and
are caused by the same source; these are referred to as \emph{``combs''}, where
the spacing is often due to some periodic effect.
Nearly all combs identified in the \ac{ASD} data begin at a certain harmonic
number and extend up to a cutoff harmonic, i.e., most combs do not span the
entire senstive frequency band, though this is not universal.
The specific details of a comb are complex and depend on factors such as the
sensitivity of the detector, the source of the noise artifact, and how the
noise couples into the detector data.

The first two aLIGO observing runs yielded highly sensitive data, enough for
detecting multiple transient gravitational wave signals from compact binary
mergers, but not yet sensitive enough for detection of persistent gravitational
wave signals.
An overview of the spectral artifacts from the first two observing runs may be
found in~\cite{Covas:2018}.
Lists of spectral artifacts are also available via 
\ac{GWOSC}~\cite{Abbott:2019ebz,gwosc_o1_strain,gwosc_o2_strain}.
In summary, the first observing run was found to have many lines and combs
while the second observing run had fewer lines and combs, though still quite
many that are problematic and many that remain
unidentified~\cite{O1_lines_lists,O2_lines_lists}.
One of the most prolific and problematic spectral artifacts was a comb of
lines, spaced 1~Hz apart, that polluted the majority of the \ac{LHO} and
\ac{LLO} spectra in \ac{O1}.
This was traced to blinking \acp{LED} in the LIGO data acquisition and timing
system electronics and was largely mitigated in \ac{O2} (though not entirely
eliminated).

The data quality for persistent gravitational wave searches in \ac{O1} and \ac{O2} is
discussed in detail in~\cite{Covas:2018}; here we describe and discuss efforts to
identify and mitigate spectral artifacts specifically in \ac{O3}.
Investigations remain ongoing to identify and mitigate sources of coupling.
Given the relative improvement of \ac{LLO} data in \ac{O3} compared to previous
runs, we are hopeful that these mitigation efforts will continue in the future.

\subsection{\label{sec:selfgating}Self-gated gravitational wave strain for glitch impact
mitigation}
Large amplitude transient glitches can impact \ac{PSD} estimation, causing
elevated noise floor levels compared to times when no glitches are present.
Prior to \ac{O3}, the probability for a high-amplitude transient glitch for a
given stretch of coherently analyzed data was relatively low.
In \ac{O3}, however, the gravitational wave strain data is subject to relatively frequent and
large amplitude glitches, such that the probability for one or more glitches
in a given coherent segment is high.
This strongly degrades analyses for persistent gravitational wave signals that
rely on an accurate measurement of the noise \ac{PSD}.
Persistent gravitational wave searches are robust against vetoes that trigger
on the loud glitches in the gravitational wave strain time series, so a scheme that
removes these glitches from the data has been implemented~\cite{self_gating}.
This is referred to as \emph{``self-gating''}, and is similar, though not
identical to, the gating used for transient gravitational wave searches (see
\sref{s:gating}).

Both approaches remove data when loud glitches occur using a smooth windowing
function around each glitch time period.
The novel aspect of this approach is: 1) the self-gating scheme removes data
when certain band-limited-root-mean-square thresholds are exceeded; and 2) the
thresholds for removal of data are much stricter so that many more large
amplitude glitches are removed from the data.
Further details can be found in~\cite{self_gating}.

\begin{figure}[t]
    \centering
    \includegraphics[width=1\textwidth]{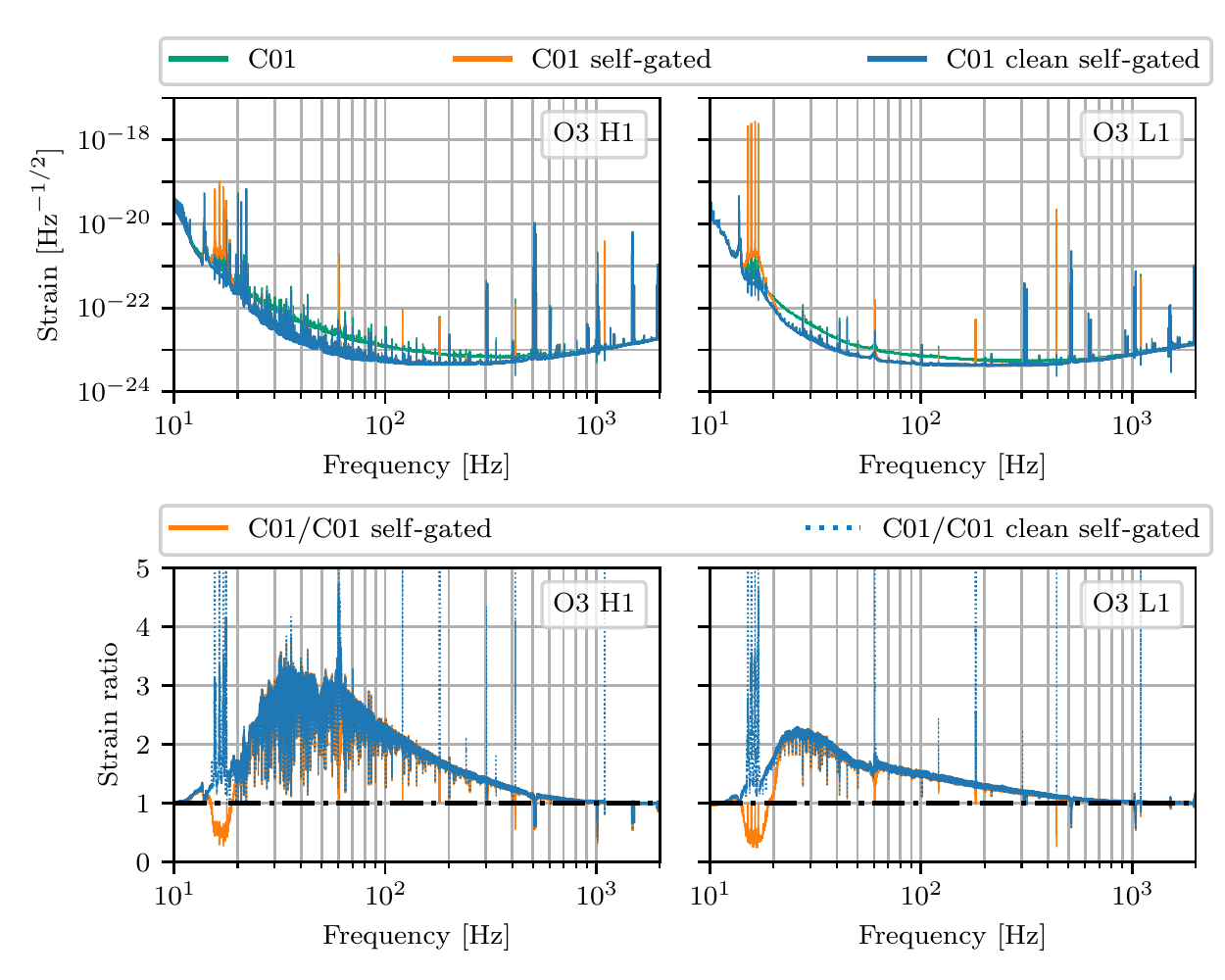}
    \caption{
    Run-averaged, noise-weighted \ac{ASD} curves for the \ac{O3} observing run.
    \textit{Top panels:} a comparison of the \ac{LHO} (left) and \ac{LLO} (right)
    \acp{ASD} for C01 data (green), C01 self-gated data (orange), and C01 cleaned,
    self-gated data (blue).
    Note the orange and blue curves are significantly below the green curve and
    some lines are cleaned.
    \textit{Bottom panels:} ratio of the \ac{LHO} (left) and \ac{LLO} (right) C01/C01
    self-gated \acp{ASD} (orange) and C01/C01 cleaned, self-gated \acp{ASD}
    (blue dashed). \ac{ASD} ratio values above 1 indicate improved sensitivity
    due to the self-gating procedure.
    The orange curve dips below 1 in a few specific regions around strong
    instrumental lines, indicating some degradation caused by the self-gating
    procedure when used on data with high-SNR calibration lines.
    }
    \label{fig:O3runaverageSpectra}
\end{figure}

Using self-gated data greatly improves the \ac{PSD} estimation, especially for
frequencies below 500~Hz.
\Fref{fig:O3runaverageSpectra} illustrates the improvements to the \ac{O3} run-
averaged, noise-weighted \ac{ASD} estimation for \ac{LHO} and \ac{LLO} strain
data.
Using the self-gating procedure mitigates the elevated noise floor by a factor
of 1.5 to 3 across a broad frequency span of 20 to 250~Hz.
Data that has not been cleaned of calibration lines and 60~Hz power mains
lines (C01 data)~\cite{VietsThesis} is degraded around those frequencies as
well as the suspension violin mode resonances when using the self-gating
procedure.
The degraded sensitivity around calibration and power mains lines is mitigated
using data that has applied all of: 1) linear subtraction of the calibration
lines and power mains harmonics; 2) cleaning of the non-stationary 60~Hz
sidebands~\cite{Vajenta:2020ml}; and 3) self-gating of the loud glitches.

\subsection{New or newly identified non-astrophysical spectral artifacts}
In each observing run, narrow spectral artifacts in full-run average amplitude
spectral density are flagged and, where possible, identified to be caused by
certain known disturbances~\cite{Covas:2018,O1_lines_lists,O2_lines_lists}.
A similar approach is also employed in \ac{O3}, where 7200-s-long Fourier
transforms are generated---covering all of \ac{O3} observing mode epochs---
averaged, and noise-weighted together in order to uncover narrow spectral
features that could impact persistent gravitational wave searches.
We have successfully identified several new features and summarize them below.
\begin{enumerate}
\item In \ac{LHO}, loud lines near 20~Hz are visible in the run-average
  spectra.
  These features are caused by sinusoidal injections into an optical alignment
  control loop in order to maintain stability of the angular control system.
  These lines could not be moved outside the LIGO \ac{LHO} detection band, as
  they were able to do in \ac{LLO}.
  Thus, they appear as strong spectral features with non-linear sidebands
  caused by mixing of the signals with low frequency (micro-)seismic noise.
\item \label{enum:item:callines} In both \ac{LHO} and \ac{LLO}, loud sinusoidal
  excitations are purposefully added to the control loop sensitive to the
  interferometer arm length differences.
  These generate fiducial length changes to the arm lengths and serve as
  calibration references (referred to as ``calibration lines'').
  Such loud excitations, however, are observed to generate non-linearities by
  the digital-to-analog converters and several other spectral artifacts are found to be
  multiples of the fundamental frequency of the calibration lines or mixing
  between calibration line frequencies, e.g., $f_1 + f_2$.
\item In both \ac{LHO} and \ac{LLO}, the calibration lines are also found to
  mix with the occasionally loud violin resonances of the suspended optics.
  Typically the violin resonances are not strongly excited, but there are
  occasional instances, following an earthquake, for example, when the
  resonances are excited.
  In these time periods, mixing between the violin resonances and calibration
  lines becomes apparent.
  Violin resonances are extremely high-Q features by design and challenging to
  control and damp out.
  They also are roughly independent, so they can be excited differently at
  different times.
  All of these features complicate mitigation of these spectral artifacts.
\item In addition to the calibration lines located at static frequencies described in
  \ref{enum:item:callines}, at both \ac{LHO} and \ac{LLO}, several additional
  calibration lines are added to the control loop sensitive to the difference
  in interferometer arm lengths \emph{sequentially} above 1~kHz.
  The actuator to add this line is not particularly strong, so the measurements
  require integrating for at least 1-day of low-noise operation.
  Once enough data is collected at a particular frequency, the line is moved to
  another frequency by 500~Hz for another 1-day period.
  After the highest frequency is reached, approximately 5~kHz, the entire process
  repeats, beginning again at 1~kHz.
  These measurements enable direct characterization of the detector performance
  at frequencies above 1~kHz and reduce uncertainty on systematic errors of the
  detector calibration~\cite{Sun_2020}.
\end{enumerate}

\subsection{Other non-astrophysical artifacts}
Several other artifacts are clearly non-astrophysical, without having a clear
source of noise that causes the spectral artifacts in the data.
We report those artifacts here along with our understanding to date.
Investigations remain ongoing into these artifacts using tools described
in~\cite{Covas:2018} and new approaches such as~\cite{Harris:2020}.
\begin{enumerate}
\item Several near 1-Hz combs have been identified in \ac{LHO} data.
  These combs are measurably different than an exact 1-Hz comb that has been
  largely mitigated (though not completely eliminated).
  The exact 1-Hz comb is caused by blinking \acp{LED} of different GPS-
  synchronized electronics components~\cite{Covas:2018}.
  It may be possible that there are un-synchronized blinking \acp{LED} in
  electronics that are causing near 1-Hz combs in the data.
\item Several 2-, 4-, 8-, and 16-Hz combs with offsets not clearly understood
  are found in \ac{LHO} data.
  These combs may be linked to the digital system, since the comb tooth spacing
  is exactly integer values (to measurement precision of 7200-s-long
  \acp{FFT}).
  The source and coupling of this comb is not yet clearly understood.
\item A near 30-Hz comb is observed only in \ac{LHO} data with a coupling
  that is not yet understood.
  This comb was initially identified as a near 60-Hz comb, with some
  speculation that there may be an unidentified electrical component with poor
  grounding.
  Recognizing that the comb spacing is instead half the original identification
  requires further study to identify the coupling mechanism.
\item Several unphysical, narrow, downward excursions in the gravitational wave strain noise
  \ac{ASD} are identified near violin resonances of the suspended test masses.
  These artifacts indicate unphysical values in the data and should therefore
  be removed from persistent gravitational wave search analyses.
  They may be caused by either an unmodelled parasitic coupling to length
  control signal around the violin resonances resulting in mis-calibrated data
  or an overdamping of violin resonances that result in control loop ``gain
  peaking'' at adjacent frequencies.
\end{enumerate}

\subsection{\label{sec:cohnoiseinvest}Investigations of coherent noise}
Searches for the \ac{GWB} are based on cross-correlating the gravitational wave strain
data between detectors~\cite{Allen:1997ad,Romano:2016dpx}.
In order to make a statement about the presence or absence of astrophysical
sources of correlation, it is crucial to understand and control noise sources
that are correlated between detectors.
Since \ac{GWB} searches integrate over an entire observing run, the most
important correlations are typically much smaller than those that affect transient
searches, requiring additional analysis.

\begin{figure}[t]
  \centering
  \includegraphics[width=\textwidth]{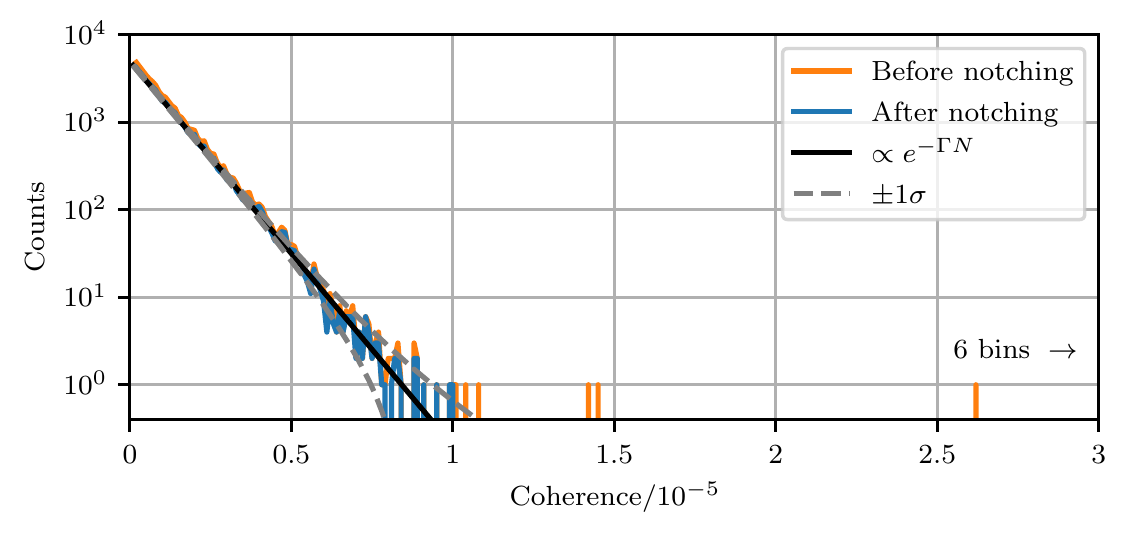}
  \caption{
  Histogram of measured coherence values per frequency bin between \ac{LHO}
  and \ac{LLO} gravitational wave strain data before (orange curve) and after (blue curve)
  removing known instrumental lines.
  The expectation for uncorrelated Gaussian noise (black curve), is consistent
  with the measured coherence once instrumental lines are removed.
  Note that 6 frequency bins have coherences larger than $3 \times 10^{-4}$ and
  do not appear in this plot.
  Five of these were notched; the sixth at 33.22 Hz was the adjacent frequency
  bin to a known calibration line non-linearity at 33.2 Hz, but the coherence
  of this frequency bin is not statistically significant in the stochastic
  search results.
}
\label{fig:O3_HL_coherence}
\end{figure}

As described in \sref{sec:coh} and~\cite{Covas:2018,Coughlin:2010zza}, the
coherence is measured between gravitational wave strain channels at different sites, and
between the gravitational wave strain channel at one site with auxiliary channels at the
same site.
Coherences between gravitational wave strain channels identify frequencies with a large amount of
correlation, which can be followed up by looking at coherences with auxiliary
channels.
Auxiliary channels not measuring the interferometer control loop sensitive to
gravitational waves should have extremely low coherence with the gravitational wave strain channel.
Frequency bins containing lines that are known to have an instrumental origin
are removed from the analysis.
\Fref{fig:O3_HL_coherence} shows a histogram of coherence values in each
frequency bin between the \ac{LHO} and \ac{LLO} strain channels observed in
\ac{O3}.
After removing instrumental lines, the distribution is consistent with the
expectation for uncorrelated Gaussian noise in
\eref{eq:coherence-distribution}.

\begin{figure}[t]
  \centering
  \includegraphics[width=\textwidth]{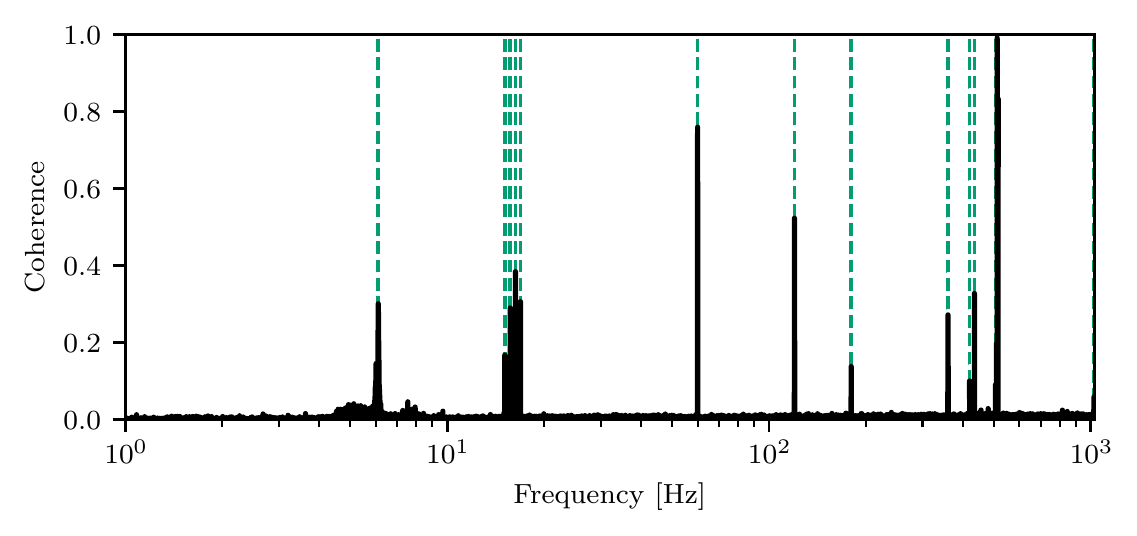}
  \caption{The measured coherence between the gravitational wave strain channel at
    \ac{LLO} and an auxiliary channel monitoring interferometer readout optics
    alignment (i.e., L1:SQZ-OMC\_TRANS\_RF3\_Q\_NORM\_DQ) (solid black curve).
    Green dashed vertical lines draw the eye to specific frequencies of
    significant coherence, in units of hertz: 6.095, 6.127, 15.1, 15.7,
    16.3, 16.9, 59.959, 119.932, 179.924, 359.831, 419.864, 434.9,
    505.718, 508.772, 511.604, 515.478, and 1023.331.
    The low-coherence values at other frequencies are not significant.}
\label{fig:coherence-tool}
\end{figure}

A specific example of a line follow-up with the coherence tool is illustrated
by \fref{fig:coherence-tool}.
This example shows the measured coherence between the \ac{LLO} gravitational wave strain
channel and a channel measuring particular angular misalignments of the
interferometer readout optics.
This allows rejection of spurious noise outliers that would otherwise
degrade persistent gravitational wave analyses.

A sub-threshold comb search is used to search for the existence of combs 
that are coherent between the Hanford and Livingston detectors in \ac{O3}.
These combs may not be visible directly in the coherence measurement but may
still impact searches.
A large number of coherence measurements are generated from many
realizations of Gaussian and uncorrelated noise.
From this simulation, a background distribution is computed in order to set a
threshold \acp{SNR} on combs that may be present in real data.
The background distribution can then be used to assign a statistical
significance for the comb-finder \ac{SNR}.
After removing known instrumental lines, we find that there is no statistically
significant evidence of sub-threshold combs impacting the \ac{GWB}
search in \ac{O3}.
We have provided the list of frequencies removed from \ac{GWB} analyses for the
\ac{O1} search in~\cite{Covas:2018}, 
for \ac{O2} in~\cite{O2StochIsoDataRelease}, 
and for \ac{O3} in~\cite{O3StochIsoDataRelease}.

\section{Future prospects}\label{s:O4expectations}

A high rate of expected detections will drive preparations for 
future observing runs.
The rate of confident gravitational wave detections is expected
to increase for the next observing run (\ac{O4}) by a factor of
4 and could be as high as
one detection per day~\cite{Aasi:2013wya}. 
At design sensitivity, the
confident detection rate could be more than one per day, plus an
even larger number of potentially astrophysically interesting marginal
events.
As the event rate grows, detector characterizations methods
will require emphasis on standardization and automation
to handle the large number of  
observed gravitational wave signals.
Concurrently, there are ongoing efforts to prepare
for the detection of gravitational waves from
new source classes.

Upgrades to the LIGO detectors in preparation for the
next observing run will introduce
new components that will require additional characterization 
before the start of \ac{O4}~\cite{Aasi:2013wya}.
The increasing duration of each observing run and expectation of high
uptime will further place demands on timely
identification of instrumental artifacts. 
To address the increase in complexity of the instruments and 
potential new sources of noise, 
statistical and machine-learning based approaches, 
which can more easily take advantage of the large number of 
data streams at each site, are expected to be used
more frequently~\cite{Smith:2011an,Walker:2018ylg,Cuoco:2020ogp}.

While the data quality methods described in prior sections were successful 
in enabling the confident detection of at least 
transient gravitational wave signals
during the first three advanced-era observing runs 
(\ac{O1}, \ac{O2}, and \ac{O3a}), 
the techniques employed were often time-consuming, even with
the use of a sophisticated computational infrastructure. 
For example, 
some data quality vetoes were tuned and tested 
for optimum performance 
by hand, a process that could take days, and validation of the first 
several detected gravitational wave events took multiple
months for each event. 
Tools based on machine learning methods~\cite{Cuoco:2020ogp} 
have shown promise in helping reduce the time and human input required
for these efforts.

Looking forward to O4, the rapid increase in detector sensitivity
will require careful planning from the LIGO detector characterization
team in order to address the large number of gravitational wave detections, 
both from compact binaries and potentially novel sources.  
Detections of new gravitational wave source types, such as those
from persistent gravitational wave sources, 
will require similar levels of development and investigation
as was completed for the first gravitational wave detection~\cite{GW150914_detchar}.
More generally, we expect that 
the LIGO data quality and detector characterization
strategy in future runs will continue to move increasingly toward 
automation and quantitative metrics both in low latency and 
for the offline analyses. 
Toward this end, we anticipate the growth and development
of novel techniques
that harness the power of the computational infrastructure
supporting LIGO data quality studies.

\section{Acknowledgments}
The authors would like to thank members of the
Virgo and KAGRA detector characterization groups
who have contributed to an environment of open 
collaboration that enabled this work.

The authors gratefully acknowledge the support of the 
United States National Science Foundation (NSF) 
for the construction and operation of the 
LIGO Laboratory and Advanced LIGO as well as the 
Science and Technology Facilities Council (STFC) 
of the United Kingdom, and the Max-Planck-Society (MPS) 
for support of the construction of Advanced LIGO. 
Additional support for Advanced LIGO was provided by the 
Australian Research Council.

LIGO was constructed by the California Institute of Technology 
and Massachusetts Institute of Technology with funding from 
the National Science Foundation, 
and operates under cooperative agreement PHY-1764464. 
Advanced LIGO was built under award PHY-0823459.
The authors are grateful for computational resources provided by the 
LIGO Laboratory and supported by 
National Science Foundation Grants PHY-0757058 and PHY-0823459.
This work carries LIGO Document number P2000495.

We would like to thank all of the essential workers who put 
their health at risk during the COVID-19 pandemic, 
without whom we would not have been able to complete this work.

\section{References}
\bibliographystyle{iopart-num}
\bibliography{o3-detchar-paper.bbl}

\end{document}